# Statistical foundations of revised reverberation theory for a diffuse sound field in a room

Toshiki Hanyu,[1,a]

[1] *Department of Architecture and Living Design, Nihon University Junior College, 7-24-1 Narashinodai, Funabashi, Chiba 274-8501, Japan*

## ABSTRACT

Revised reverberation theory is reformulated statistically for a diffuse sound field to clarify how the mean residual free path connects direct sound, reflected sound, and room energy decay. Under completely absorptive conditions, direct sound persists until reaching a boundary, producing decay after the source stops even without reflections. An integral-geometric relation connects the mean residual free path to the second moment of the wall-to-wall free path, thereby linking the mean propagation distance of direct sound to the temporal variance of reflected sound under an independent-renewal approximation. Using this relationship, the macroscopic decay rate and the microscopic long-time asymptotic decay rate agree exactly through second order in the absorption coefficient under an independent-renewal approximation. A gamma model with a consistent residual free-path distribution eliminates probability at negative arrival times. Ray-tracing simulations in three rectangular rooms support the predicted initial decay slopes under completely absorptive conditions and reverberation predictions for absorption coefficients up to 0.5. Engineering approximations yield macroscopic formulas requiring only room volume, surface area, and absorption coefficient while reducing prediction errors relative to the Sabine and Eyring formulas. A supplementary fixed-source and local-receiver test supports T30 prediction within the tested conditions. These results establish a common statistical basis for direct and reflected sound and connect the physical interpretation of reverberation to practical prediction.

[a] Email: hanyu.toshiki@nihon-u.ac.jp, toshiki.hanyu@gmail.com

## I. INTRODUCTION

The Sabine and Eyring formulas have provided a foundation for room acoustics by relating reverberation time to room volume, surface area, and absorption coefficient [Sabine1922; Eyring1930]. Their significance lies in both computational simplicity and the physical relationship between stored acoustic energy and boundary absorption. Different assumptions concerning diffuseness, reflection intervals, and energy decay nevertheless lead to different formulas [Stephenson2012]. A reverberation theory must therefore be assessed by the process it represents as well as by its prediction accuracy.

Free-path variability has been incorporated into statistical descriptions of reverberation. Kuttruff derives a correction using independent wall-to-wall free paths and relates their variance to that of the reflection count [Kuttruff2024] (Sec. 5.5, pp. 120–123). Polack examined reflection histories and revisited Kuttruff's gamma model of flight times between boundary reflections [Polack1992; Polack2025]. These studies establish how free-path variance affects decay beyond the information supplied by the mean alone.

Geometry-specific application requires additional statistics, such as free-path variance. When these must be calculated numerically, the practical advantage of a compact formula becomes less clear. The challenge is to connect their physical meaning to predictions using variables readily available in acoustic design.

Complete absorption raises a related physical question. Sabine's formula gives a finite reverberation time at unit absorption, whereas Eyring's time approaches zero. This difference concerns how sound energy decays after the source stops when no reflected sound exists. Previous statistical studies have examined this limit [Polack2025]; revised reverberation theory addresses it through the physical concept of reverberation of a direct sound [Hanyu2024; Hanyu2025; arXiv_V1].

Sound already propagating in a room remains unaffected by boundary absorption until it reaches the next wall. Even with completely absorptive walls, its energy therefore does not disappear instantaneously when the source stops. The resulting direct-sound decay extends the description of reverberation beyond multiple reflections.

The author introduced the mean free path of direct sound, $l_d$, and derived a macroscopic formula from room energy balance using the geometric-cutoff approximation $l_d = l(1-1/e)$, where $l$ is the mean wall-to-wall free path [Hanyu2024]. A subsequent microscopic formulation described sound at successive reflection orders [Hanyu2025], and the relationship between the two models was examined in [arXiv_V1]. Because $l_d$ also enters the reflected-sound process, that study adopted the more general term "mean residual free path." The aim is to reconstruct the physical basis of reverberation formulas around this quantity, rather than merely add a correction to a classical formula.

Three issues remained. The earlier microscopic models assumed $(l_d/c)^2$ for the temporal variance without deriving it from the statistics associated with $l_d$, where $c$ is the sound speed [Hanyu2025; arXiv_V1]. The macroscopic and microscopic decay-rate expansions agreed at first order in absorption, but only approximately at second order [arXiv_V1]. Finally, the normal approximation assigned probability to negative arrival times.

This study gives the revised theory a statistical foundation under a diffuse-field idealization. An established integral-geometric relation connects the mean residual free path $l_d$ to the mean wall-to-wall free path $l$ through the second moment [Mazzolo2008]. Together, these lengths determine the variance of an individual propagation interval and, under an independent-renewal approximation, the temporal variance expansion of successive wall-collision times. Thus, a quantity introduced through direct sound also characterizes propagation after reflection.

Using common free-path statistics, we establish exact agreement through second order between the macroscopic decay rate and the microscopic long-time rate, and interpret the macroscopic model through mean energy lifetime. A gamma free-path approximation and its consistent residual distribution exclude negative arrival times. Collision-time densities are distinguished from the collision-count probabilities that determine the energy remaining in the room. Variance corrections and gamma models have precedents; the contribution here is their connection through the mean residual free path within a common description of direct sound, reflected sound, and total energy decay.

Diffuse reflection does not ensure independent free paths or a uniformly mixed surviving field. We therefore distinguish the general correlated process from the independent-renewal, distributional, and engineering approximations. Mean lifetime, long-time decay rate, and the 60-dB crossing of the decay curve are treated as distinct quantities.

Ray tracing in three rectangular rooms tests predictions using either geometry-derived statistics or fixed residual-to-mean free-path ratios. The latter permit formulas based on room volume, surface area, and absorption coefficient. Prediction accuracy is assessed over the practical range defined here as $\alpha \le 0.5$; complete absorption is used to examine direct-sound decay and the physical meaning of the residual free path.

A supplementary test in a separate rectangular room examines fixed omnidirectional sources and local receivers. It compares T30 with the spatially uniform initial-condition reference and evaluates engineering prediction errors without changing the source-position-averaged theoretical target.

Sections II and III define the stochastic framework and derive the residual free-path relations. Sections IV and V develop the decay model and establish its macroscopic correspondence. Section VI introduces the gamma model and engineering approximations, Sec. VII presents numerical verification, and Sec. VIII gives the conclusions.

## II. PHYSICAL AND STOCHASTIC FRAMEWORK

### A. Diffuse sound field and modeling assumptions

A perfectly diffuse sound field provides the idealized reference: energy is spatially uniform and propagation directions are isotropic [Kuttruff2024]. Diffuse wall scattering, however, does not ensure that these conditions persist throughout decay.

Sabine's theory assumes continued uniformity and isotropy, making the energy flux incident on walls proportional to the spatially averaged acoustic energy density [Sabine1922; Kuttruff2024]. Here, particles propagate at finite speed and lose energy only at wall encounters. Decay therefore depends on free-path statistics and collision histories [Kuttruff2024; Polack2025].

The source position is sampled uniformly within the room and its initial emission direction isotropically. The theory describes an expectation over these variables and subsequent propagation, rather than the deterministic response to a fixed source. This averaging can represent repeated realizations of a single source; it does not require simultaneous sources.

Sound energy is represented by infinitesimal energy elements, hereafter referred to as sound particles, that travel along straight paths at the sound speed c. Each sound particle propagates freely until it reaches a wall, where it is either absorbed or reflected. Conditional on the collision position and the local wall normal, the reflected direction is sampled according to Lambert's cosine law, and its conditional distribution is assumed to be independent of the incident direction [Comets2009]. Nevertheless, because the reflection position is determined by the preceding propagation path and the room geometry, successive wall-to-wall free paths are not necessarily statistically independent.

For wall absorption, the basic model initially assumes a uniform absorption coefficient α over all wall surfaces. Absorption decisions at successive wall encounters are assumed to be independent of the geometrical propagation path and of previous absorption decisions. At each encounter, a sound particle is absorbed with probability α or reflected with probability 1 − α. In terms of expected sound energy, this probabilistic description of absorption is equivalent to multiplying the energy by the factor 1 − α at each reflection. If the absorption coefficients differ among wall surfaces, the absorption probability depends on the location of each wall encounter, and replacing these coefficients by an average absorption coefficient constitutes an additional approximation that must be considered separately.

The general trajectory model retains correlations. The independent-renewal approximation is introduced for analytical treatment, followed by distributional and engineering approximations where needed.

In this study, reverberation is defined as the decay of the total sound energy remaining in a room, including both direct and reflected sound, after the sound source stops emitting sound following the establishment of a steady state. This definition follows the concept of the reverberation of direct sound developed in the revised reverberation theory [Hanyu2024; Hanyu2025; arXiv_V1]. The present description requires a finite steady-state energy. When only wall absorption is considered and air absorption is neglected, the lossless case α = 0 is excluded from the decay response after the steady source stops and is considered, where needed, as a limiting case. Reverberation time is defined as the time required for the spatially averaged acoustic energy density to decrease by 60 dB relative to its value when the source stops. Assuming an exponential decay curve is distinguished from this physical definition and is treated as an approximation in the macroscopic model discussed later.

### B. Wall-to-wall and residual free paths

To describe reverberation in stochastic terms, the free paths of sound particles are distinguished into wall-to-wall free paths and residual free paths. The former is the distance traveled by a sound particle after reflection at a wall until it reaches the next wall, whereas the latter is the distance traveled by a sound particle from an arbitrary position within the room until it reaches the first wall.

Let V and S denote the room volume and total wall surface area, respectively. The wall-to-wall free path is represented by a random variable L, with probability density function $p_L(s)$. The starting position on the wall is sampled with probability proportional to surface area, and the propagation direction is selected according to Lambert's cosine law. For free flights to the next boundary within a closed, bounded room, with S denoting the area of the entire boundary enclosing the domain, these sampling conditions give the following mean, irrespective of the detailed room shape [Kosten1960; Mazzolo2003]:

$$l = E[L] = \frac{4V}{S} \qquad \text{(II-B1)}$$

Here, l is the mean free path. This relation represents the expectation under the specified probabilistic conditions; it does not imply that the mean free path at every reflection order immediately equals l for sound particles launched under arbitrary initial conditions.

In contrast, in accordance with the stochastic model defined in Sec. II A, the source position is treated as a random variable uniformly distributed throughout the room, and the initial emission direction is selected isotropically. The distance from the emission position to the first wall encounter is defined as the residual free path R. Denoting its probability density function by $p_R(r)$ and its expectation by $l_d$, we obtain

$$l_d = E[R] \qquad \text{(II-B2)}$$

The quantity $l_d$ is the mean residual free path, which has been used as a fundamental quantity for describing the reverberation of direct sound in the revised reverberation theory [Hanyu2024; Hanyu2025; arXiv_V1]. Because R corresponds to the distance traveled by a sound particle before it experiences any wall reflection, it is related to the stochastic description of direct sound at reflection order zero.

The two paths use different joint sampling measures: surface-area weighting with Lambertian directions for L, and uniform volume weighting with isotropic directions for R. Their distributions and means therefore generally differ. Lambert's conditional angular law relative to a local normal does not itself imply anisotropy of the room-wide field.

The mean residual free path is related to the second wall-to-wall free-path moment [Mazzolo2008]; Sec. III derives this connection.

Successive wall-to-wall paths can be correlated through collision positions and room geometry. Their reflection-order distributions can also differ from the stationary wall-collision density because of the initial state. Marginal statistics must therefore be distinguished from the joint distribution; independence is introduced only as a subsequent approximation.

These are geometric statistics before absorption weighting. The distribution of particles surviving at a given time can differ, as described by the collision-count formulation below.

### C. Reflection counts and stochastic energy survival

The free paths defined in the preceding section are used to relate the history of wall collisions to the fraction of acoustic energy remaining in the room. Descriptions of reverberant decay based on reflection-count statistics have been considered in previous studies [Polack2025]. Here, the initial residual free path is distinguished from the subsequent wall-to-wall free paths, and correlations between reflections are retained in the formulation. This section considers wall absorption only; air absorption is excluded.

First, consider a geometric sound-particle trajectory that continues to undergo wall reflections without being terminated by absorption. Let $t=0$ denote the emission time of the sound particle, and let $T_n$ denote the elapsed time to its $n$th wall collision. For a constant sound speed $c$,

$$T_n=\frac{1}{c}\left(R+\sum_{j=1}^{n-1} L_j\right), n=1,2,\ldots \quad \text{(II-C1)}$$

where $R$ is the residual free path to the first wall collision and $L_j$ is the free path between the $j$th and $(j+1)$th wall collisions. For $n=1$, the sum is taken to be zero, so that $T_1=R/c$. The number of wall collisions up to time $t$ is defined as

$$N(t)=\sum_{n=1}^{\infty} 1\{T_n\le t\}. \quad \text{(II-C2)}$$

Here, $1\{\cdot\}$ is the indicator function, equal to unity when the stated condition holds and zero otherwise. The number of collisions in any finite time interval is assumed to be finite. The statistical dependence among $R, L_1, L_2, \ldots$ is incorporated in the distributions of the collision times and $N(t)$; independence of the free paths is not required for this definition.

Let $\alpha$ be the energy absorption coefficient common to all walls, and let $q=1-\alpha$ be the fraction of energy retained upon reflection. Under the assumption in Sec. II A that the absorption event at each collision is independent of the geometric trajectory and of previous absorption outcomes, the survival probability of a sound particle along a trajectory with $N(t)=n$ is $q^n$. The initial acoustic energy introduced at $t=0$ is represented by an ensemble of sound particles, each carrying the same initial energy. Because each particle retains its energy until it is absorbed, the mean fraction of surviving particles equals the mean fraction of the initial acoustic energy remaining in the room. Denoting this mean energy survival fraction by $F(t)$, we obtain

$$F(t)=E\left[q^{N(t)}\right]=\sum_{n=0}^{\infty} q^n Pr\{N(t)=n\}. \qquad \text{(II-C3)}$$

The expectation is taken over the propagation process, including the source position, initial emission direction, and reflection directions. The function $F(t)$ is dimensionless and satisfies $F(0)=1$ and $0\le F(t)\le 1$. Here, $Pr\{N(t)=n\}$ is the fraction of trajectories that have undergone exactly $n$ collisions by time $t$; it is distinct from the probability density of the $n$th wall-collision time $T_n$.

The component that has not yet reached a wall corresponds to $N(t)=0$, with probability $Pr\{R>ct\}$. Under completely absorptive conditions, $\alpha=1$, only this component remains, giving $F(t)=Pr\{R>ct\}$. In this case, the weight of the $n=0$ term in Eq. (II-C3) is defined to be unity. Thus, the formulation preserves the physical property that direct-sound energy remains in the room until the first wall collision, even when no reflected sound exists [Hanyu2024; Hanyu2025; arXiv_V1].

Next, the mean survival fraction $F(t)$ of the initial acoustic energy is related to the decay curve after a source operating at constant acoustic power stops emitting sound. Let $W$ [J/s] denote the acoustic power radiated by the source. A source at a fixed position in the room is operated at constant acoustic power until a steady state is reached and then stops emitting sound at $t=0$. The resulting response is averaged over the uniform distribution of source positions specified in Sec. II A. In the following, $t$ denotes the elapsed time after the source stops. Assuming that the energy response is linear and time invariant, the average acoustic energy density $E(t)$ after the source stops is obtained by superposing the contributions of the energy emitted before the source stops:

$$E(t)=\frac{W}{V}\int_t^{\infty} F(u)\,du. \qquad \text{(II-C4)}$$

Here, $V$ is the room volume, and $W>0$ and integrability of $F(t)$ are assumed. This relation, which gives the decay curve after the source stops by backward integration, corresponds to Schroeder's integration method [Schroeder1965]. The integrand in Eq. (II-C4), however, is the mean survival fraction of the acoustic energy in the room defined in the present model, rather than the squared sound-pressure impulse response at a receiver.

Normalization by the value when the source stops gives

$$G(t)=\frac{E(t)}{E(0)}=\frac{\int_t^{\infty} F(u)\,du}{\int_0^{\infty} F(u)\,du}. \qquad \text{(II-C5)}$$

The reverberation time $T_{60}$ considered in this study is defined as the time satisfying

$$G(T_{60})=10^{-6}. \qquad \text{(II-C6)}$$

In general, the mean survival fraction $F(t)$ of the initial acoustic energy and the normalized decay curve $G(t)$ after the source stops do not have the same functional form. Equations (II-C3)–(II-C6) do not assume exponential decay and retain correlations among the free paths. The next section examines the relationship between stationary geometric statistics and the initial conditions and specifies the additional conditions required for the independent-renewal approximation.

### D. Stationary initial conditions and the renewal approximation

Stationarity here refers to an unchanged distribution of positions and directions in the geometric propagation process without absorption, not to an acoustic steady state maintained by balancing source power and losses.

For Lambertian reflections, uniform positions and isotropic directions form a stationary geometric distribution [Comets2009]. The initial ensemble in Sec. II A has this distribution after averaging over source positions; it is not the immediate field of a single fixed source.

Sampling at wall collisions yields $L$, whereas observation at an arbitrary time preferentially samples longer segments because traversal takes longer. The remaining distance is $R$. Section III derives the distributional and moment relationships produced by these different sampling measures.

Stationarity does not imply independent intervals: a collision position retains information about the preceding path. Nor does it identify the distribution at every reflection order counted from an interior start with the stationary wall-collision distribution.

When the independent-renewal approximation is introduced in the subsequent analysis, the free paths $L_1, L_2, \ldots$ following the initial residual free path $R$ are assumed to be independent random variables with a common probability density $p_L$, and are further assumed to be independent of $R$. Denoting the probability density of $R$ by $p_R$, the joint probability density under this approximation is

$$p_{R, L_1, \ldots, L_n}(r, x_1, \ldots, x_n) = p_R(r) \prod_{j=1}^{n} p_L(x_j). \tag{II-D1}$$

Here, $p_L$ and $p_R$ are the stationary wall-to-wall and residual densities. Equation (II-D1) neglects their joint dependence and enables the sequential convolutions in Sec. IV.

Absorption weights trajectories by $q^{N(t)}$, so surviving particles need not remain uniformly and isotropically distributed even when the unweighted process is stationary.

## III. RESIDUAL FREE-PATH STATISTICS AND SECOND-MOMENT RELATIONS

### A. Stationary residual free-path distribution

This section relates the statistics of free paths sampled at wall collisions to those of the residual distance from a position along a trajectory to the next wall. As defined in Sec. II B, the starting position on the wall is sampled in proportion to surface area, and the propagation direction into the room is sampled according to Lambert's cosine law. Denoting the probability density of the wall-to-wall free path obtained by this sampling procedure by $p_L(x)$, its mean is

$$l = E[L] = \int_0^{\infty} x\, p_L(x)\, dx = \frac{4V}{S}. \tag{III-A1}$$

Here, $V$ is the room volume, and $S$ is the total area of the boundary enclosing the propagation domain [Kosten1960; Mazzolo2003]. This corresponds to the mean chord-length relation based on the uniform measure on lines in integral geometry. For a nonconvex domain, multiple interior segments along the same straight line are counted separately as segments between consecutive boundary intersections. This mean is taken over the ensemble of geometric trajectories before weighting by absorption and does not require independence of successive free paths.

The relationship between the distributions of paths starting at a wall and propagating into the room and those starting uniformly within the room in isotropically distributed directions corresponds to the relationship between path-length distributions for boundary incidence and internal emission in transport theory [Mazzolo2008]. Below, the relation for straight-line propagation is derived using the sampling conditions and notation of the present study.

Consider a sound particle observed at an arbitrary time in the stationary geometric propagation process described in Sec. II D. Sampling at wall collisions counts each free path once, whereas observation during propagation introduces a weight proportional to the time required to traverse that free path. Because the sound speed is constant, this time is proportional to the free-path length. Thus, if $\tilde{L}$ denotes the full length of the wall-to-wall segment being traversed at the observation time, its probability density is

$$p_{\tilde{L}}(x) = \frac{x\, p_L(x)}{l}. \qquad \text{(III-A2)}$$

The denominator $l$ is the mean free path defined in Eq. (III-A1). Since $\int_0^\infty x\, p_L(x)\, dx = l$, the probability density in Eq. (III-A2) is normalized. The variable $\tilde{L}$ denotes the full length of the wall-to-wall segment containing the observation position, rather than the residual free path itself.

Conditional on the full segment length $\tilde{L} = x$, the observation position of a particle propagating at constant speed is uniformly distributed along that segment. Consequently, the residual free path $R$ from the observation position to the next wall is also conditionally uniform on $0 < r < x$:

$$p_{R \mid \tilde{L}}(r \mid x) = \frac{1}{x}, 0 < r < x. \qquad \text{(III-A3)}$$

Weighting this conditional density by Eq. (III-A2) and integrating yields

$$p_R(r) = \int_r^\infty \frac{1}{x} \frac{x\, p_L(x)}{l}\, dx = \frac{1}{l} \int_r^\infty p_L(x)\, dx, r \geq 0. \qquad \text{(III-A4)}$$

This is the relation between path-length distributions given in Eq. (6) of Mazzolo [Mazzolo2008], applied here to wall-to-wall and residual free paths under straight-line propagation.

Defining the tail probability of the wall-to-wall free path as $\overline{F}_L(r) = Pr\{L > r\}$, Eq. (III-A4) can be written as

$$p_R(r) = \frac{\overline{F}_L(r)}{l}. \qquad \text{(III-A5)}$$

Here, the tail probability is the probability that a free path exceeds a specified length. The overbar and subscript distinguish it from the mean acoustic energy survival fraction $F(t)$ defined in Sec. II C. Similarly, defining the tail probability of the residual free path as $\overline{F}_R(r) = Pr\{R > r\}$, we obtain

$$\overline{F}_R(r) = \int_r^\infty p_R(x)\, dx = \frac{1}{l} \int_r^\infty \overline{F}_L(x)\, dx. \qquad \text{(III-A6)}$$

The derivation uses stationarity of the geometric propagation process and a constant propagation speed, but does not require independence of successive free paths.

Under the source-position and direction averaging in Sec. II, $p_R$ in Eq. (III-A5) is the density of the distance traveled by direct sound before its first collision. This need not hold for an arbitrary fixed source.

### B. Relationship between mean residual free path and the second moment

The corresponding moment relation in Eq. (1) of Mazzolo [Mazzolo2008] gives a statistical definition of $l_d$. For straight-line propagation, it follows as below.

The mean residual free path is defined as the expectation of the residual free path $R$:

$$l_d = E[R] = \int_0^{\infty} r\, p_R(r)\, dr\,. \tag{III-B1}$$

Substituting Eq. (III-A4) and interchanging the order of integration gives

$$l_d = \frac{1}{l}\int_0^{\infty} r\left[\int_r^{\infty} p_L(x)\, dx\right] dr = \frac{1}{l}\int_0^{\infty} p_L(x)\left[\int_0^{x} r\, dr\right] dx\,. \tag{III-B2}$$

Thus, when the second moment of the wall-to-wall free path is finite,

$$l_d = \frac{1}{2l}\int_0^{\infty} x^2 p_L(x)\, dx = \frac{E[L^2]}{2E[L]}\,. \tag{III-B3}$$

This relation follows from the stationary sampling conditions described in the preceding section and does not require independence of successive free paths. Nor does it assume a particular free-path distribution, such as an exponential or gamma distribution.

The physical meaning of Eq. (III-B3) can be understood through the length weighting of segments sampled during propagation. From Eq. (III-A2), the mean full length $\tilde{L}$ of the wall-to-wall segment containing the observation position is

$$E[\tilde{L}] = \frac{E[L^2]}{l} = 2l_d\,. \tag{III-B4}$$

Conditional on the full segment length, the observation position is uniformly distributed along the segment, so the mean residual distance to the next wall is half that length. However, observation during propagation favors longer segments compared with sampling at wall collisions. Consequently, the mean residual free path is generally not simply half the mean wall-to-wall free path, $l/2$.

Thus, $l_d$, introduced through direct-sound reverberation [Hanyu2024; Hanyu2025; arXiv_V1], is fixed by the second wall-to-wall free-path moment. At speed $c$, $l_d/c$ is the mean time to the first collision under the source averaging in Sec. II A, not the mean lifetime including subsequent reflections.

Although $l$ is fixed by $4V/S$, its second moment is not. Rooms with the same mean free path can therefore have different residual means; $l_d/l$ is retained as a geometry-dependent statistic.

### C. Free-path variance and its physical interpretation

The relation between the mean residual free path and the second moment established in the preceding section can be expressed in terms of the variance of the wall-to-wall free path. Defining $\sigma_L^2 = Var(L)$ for the stationary free-path distribution sampled at wall collisions, Eq. (III-B3) gives

$$\sigma_L^2 = E[L^2] - l^2 = 2l\,l_d - l^2\,. \tag{III-C1}$$

Defining the variance normalized by the squared mean free path as

$$v = \frac{\sigma_L^2}{l^2}, \tag{III-C2}$$

we obtain

$$v = 2\frac{l_d}{l} - 1, \frac{l_d}{l} = \frac{1+v}{2}. \tag{III-C3}$$

The dimensionless quantity $v$ is the squared coefficient of variation of the free path. Equation (III-C3) expresses the established moment relation cited in the preceding section [Mazzolo2008] in terms of the relative spread of the free-path distribution.

According to Eq. (III-C3), for a given mean free path $l$, a larger free-path variance corresponds to a larger mean residual free path $l_d$. This reflects the preferential sampling of longer free paths when particles are observed during propagation. Furthermore, the nonnegativity of the variance implies

$$l_d \geq \frac{l}{2}. \tag{III-C4}$$

Equality holds only when the free path takes the constant value $l$ with probability one. Even in this case, the residual distance from the observation position to the next wall varies within the segment; the residual free path itself is therefore not constant.

An exponential wall-to-wall distribution with mean $l$ gives $v=1$ and $l_d = l$. The converse does not hold: specifying $l$ and $l_d$ fixes two moments, not a unique distribution. Their connection to classical formulas is examined in Sec. V.

Equation (III-C3) connects variance-based decay descriptions [Kuttruff2024; Polack2025] to $l_d$, introduced through direct-sound reverberation. The latter retains its physical definition as a mean distance, distinct from a variance with dimensions of length squared.

The variance of individual free paths must be distinguished from correlations between successive free paths. Equation (III-C1) determines the variance of a single free-path distribution. In contrast, the variance of the sum of multiple free paths is generally

$$Var\left(\sum_{j=1}^{n} L_j\right) = \sum_{j=1}^{n} Var\left(L_j\right) + 2\sum_{1\leq i<j\leq n} Cov\left(L_i, L_j\right). \tag{III-C5}$$

Thus, specifying $l$ and $l_d$ does not determine the covariances between free paths at different reflection orders. Moreover, as discussed in Sec. II, the free-path distribution at each reflection order counted from an initial position inside the room need not coincide with the stationary distribution sampled at wall collisions. Under the independent-renewal approximation introduced in Sec. II D, each $L_j$ follows the common distribution $p_L$, and the covariance terms vanish, reducing Eq. (III-C5) to $n\sigma_L^2$.

## IV. STOCHASTIC MODEL OF REVERBERATION DECAY

### A. Sound propagation process and delayed-renewal approximation

We now use the free-path statistics of Sec. III to evaluate the collision-history representation in Sec. II, first retaining correlations and then applying the independent-renewal approximation. Air absorption remains excluded.

Along a geometrical propagation trajectory that is continued without termination by wall absorption, let $X_n$ denote the total distance traveled up to the $n$th wall collision. Its relationship to the collision time $T_n$ defined in Sec. II C is

$$X_n = R + \sum_{j=1}^{n-1} L_j, T_n = \frac{X_n}{c}, n \geq 1. \quad \text{(IV-A1)}$$

For $n=1$, the sum is taken as zero, giving $X_1 = R$. The number of wall collisions up to time $t$ is then expressed as

$$N(t) = \sum_{n=1}^{\infty} 1_{\{X_n \leq ct\}}. \quad \text{(IV-A2)}$$

The number of collisions within any finite time interval is assumed to be finite. Equations (IV-A1) and (IV-A2) hold for each trajectory and do not require independence among $R, L_1, L_2, \ldots$.

For a uniform wall energy absorption coefficient $\alpha$ and a reflection survival factor $q = 1 - \alpha$, the mean acoustic energy survival fraction defined in Sec. II C is

$$F(t) = E\left[q^{N(t)}\right] = \sum_{n=0}^{\infty} q^n Pr\{N(t) = n\}. \quad \text{(IV-A3)}$$

Dependence enters through the cumulative distances $X_n$ and collision count $N(t)$; their distributions are not generally determined by the individual free-path marginals.

Under Eq. (II-D1), $L_1, L_2, \ldots$ are independent with common stationary density $p_L$, and independent of $R$. The initial residual density is

$$p_R(r) = \frac{\overline{F}_L(r)}{l}, \overline{F}_L(r) = Pr\{L > r\}, l = E[L]. \quad \text{(IV-A4)}$$

Because the distance to the first collision follows $p_R$, whereas subsequent intercollision distances follow $p_L$, the collision process constructed from Eq. (IV-A1) is represented as a delayed renewal process, in which the first waiting time has a different distribution from subsequent waiting times. Here, “delayed” is used in its probabilistic sense to distinguish the time to the first collision from subsequent intercollision times; it does not imply an additional delay before propagation begins.

This approximation neglects geometrical dependence but preserves the stationary relation $l_d = E\left[L^2\right]/(2l)$. Consequently, $l_d$ and the wall-to-wall variance are not specified independently.

### B. Reflection-order wall-collision time distributions

Under the independent-renewal approximation, collision-time densities follow by sequential convolution, as in the earlier microscopic model [arXiv_V1]. Here, the initial and subsequent intervals use the consistent residual and wall-to-wall distributions from Sec. III.

Let $\psi_R(t)$ and $\psi_L(t)$ denote the probability densities of the residual free-path propagation time $R/c$ and the wall-to-wall free-path propagation time $L/c$, respectively. For a constant sound speed $c$, the change of variable from distance to time gives

$$\psi_R(t) = c\, p_R(ct), \psi_L(t) = c\, p_L(ct), t \geq 0. \quad \text{(IV-B1)}$$

Both densities have units of inverse time, and each integrates to unity. Substituting the stationary residual free-path relationship $p_R(r)=\overline{F}_L(r)/l$, given in Eq. (III-A5), into Eq. (IV-B1) yields

$$\psi_R(t)=\frac{c}{l}\overline{F}_L(ct). \tag{IV-B2}$$

Thus, the distribution of the time to the first wall collision is determined by the wall-to-wall propagation-time distribution; the two distributions are not specified independently.

Let $f_n(t)$ denote the probability density of the time $T_n$ of the $n$th wall collision. Since $T_1=R/c$, the first-collision density is $f_1(t)=\psi_R(t)$. Under the independent-renewal approximation, the subsequent wall-to-wall propagation time $L_n/c$ is independent of the preceding collision time $T_n$. The probability density of $T_{n+1}=T_n+L_n/c$ is therefore

$$f_1(t)=\psi_R(t), f_{n+1}(t)=\int_0^t f_n(\tau)\psi_L(t-\tau)d\tau, n\geq 1. \tag{IV-B3}$$

Repeated application gives

$$f_n(t)=\left(\psi_R*\psi_L^{*(n-1)}\right)(t), n\geq 1. \tag{IV-B4}$$

Here, * denotes convolution with respect to time, and $\psi_L^{*(n-1)}$ denotes the $(n-1)$-fold convolution of $\psi_L$. For $n=1$, Eq. (IV-B4) is interpreted as $f_1=\psi_R$. Equations (IV-B3) and (IV-B4) follow from the assumed independence of the free paths and cannot be applied directly to a general propagation process with correlations.

We next relate the collision-time distributions to the distribution of the number of collisions up to time $t$. Let $Q_n(t)$ denote the probability that the $n$th wall collision has occurred by time $t$. Then

$$Q_n(t)=Pr\{T_n\leq t\}=\int_0^t f_n(\tau)d\tau, n\geq 1. \tag{IV-B5}$$

The occurrence of the $n$th collision by time $t$ differs from the condition that exactly $n$ collisions have occurred by that time. The latter also requires that the $(n+1)$th collision has not yet occurred. Since the collision times are ordered,

$$\begin{aligned} Pr\{N(t)=0\} &= 1-Q_1(t)=\overline{F}_R(ct), \\ Pr\{N(t)=n\} &= Q_n(t)-Q_{n+1}(t), n\geq 1. \end{aligned} \tag{IV-B6}$$

Here, $\overline{F}_R(r)=Pr\{R>r\}$ is the survival function of the residual free path. Equation (IV-B6) follows from the relationship between collision times and the collision count and does not require independence of the free paths. It also holds for a general correlated process when $Q_n$ denotes the actual cumulative distribution of the corresponding collision time. By contrast, obtaining $Q_n$ from the convolutions in Eqs. (IV-B3) and (IV-B4) requires the independent-renewal approximation.

The density $f_n(t)$ describes the arrival time at the $n$th wall collision; it is not itself the acoustic energy of the $n$th-order reflected sound remaining in the room at time $t$. To obtain that energy, the propagation state between the $n$th and $(n+1)$th collisions must be combined with the survival factor $q^n$ due to wall absorption. In the next subsection, Eq. (IV-B6) is used to separate the direct and reflected sound energies and to express their decay curves after a source operating at constant acoustic power stops emitting sound.

### C. Decomposition of direct and reflected sound energy

In the sound-particle model specified in Sec. II A, direct and reflected sound are distinguished by their wall-collision histories. Let $F_d(t)$ denote the fraction of the acoustic energy supplied at time zero that has undergone no wall collision by time $t$, and let $F_r(t)$ denote the surviving fraction that has undergone at least one collision. The mean survival fraction $F(t)$ of the total energy defined in Sec. II C is decomposed as

$$\begin{aligned} F_d(t) &= Pr\{N(t)=0\} = \overline{F}_R(ct), \\ F_r(t) &= \sum_{n=1}^{\infty} q^n Pr\{N(t)=n\}, \\ F(t) &= F_d(t)+F_r(t). \end{aligned} \qquad \text{(IV-C1)}$$

Because direct sound is the component that has not yet experienced wall absorption, $F_d(t)$ is independent of the wall absorption coefficient. This persistence of energy before its first wall collision provides the basis for direct-sound reverberation [Hanyu2024; Hanyu2025; arXiv_V1].

Let $F_n(t)$ denote the survival fraction of the $n$th-order reflected sound. Equation (IV-B6) gives

$$F_n(t) = q^n\left[Q_n(t) - Q_{n+1}(t)\right], n \geq 1, \qquad \text{(IV-C2)}$$

with $F_r(t) = \sum_{n=1}^{\infty} F_n(t)$. Each component is nonnegative, and the classification by collision count involves no overlap. This decomposition does not require independence of the free paths. When the independent-renewal approximation is adopted, the sequential convolutions in the preceding subsection are used to evaluate $Q_n$.

Next, consider a source that emits at constant acoustic power $W$ until a steady state is reached and then stops emitting sound at time zero, under the averaging conditions for source position and initial emission direction specified in Sec. II A. Applying Eq. (II-C4) of Sec. II C to each component gives the mean acoustic energy densities in the room after the source stops:

$$\begin{aligned} E_d(t) &= \frac{W}{V}\int_t^{\infty} F_d(\tau)\,d\tau, \\ E_r(t) &= \frac{W}{V}\int_t^{\infty} F_r(\tau)\,d\tau, \\ E(t) &= E_d(t)+E_r(t). \end{aligned} \qquad \text{(IV-C3)}$$

The integrals are assumed to be finite. Because each component is classified by its collision history up to the observation time, energy classified as direct sound when the source stops is included in the reflected component if it subsequently undergoes a wall reflection.

Substituting Eq. (IV-C1) into Eq. (IV-C3) and changing to the distance variable $x = c\tau$ yields

$$E_d(t) = \frac{W}{Vc}\int_{ct}^{\infty} \overline{F}_R(x)\,dx, E_d(0) = \frac{W\,l_d}{Vc}. \qquad \text{(IV-C4)}$$

The value when the source stops recovers the earlier direct-sound energy relation [Hanyu2024]. The impulse survival fraction $F_d(t)$ and the energy after the source stops $E_d(t)$ remain distinct, the latter being its backward integral.

### D. Direct-sound decay under complete wall absorption

Equation (IV-C4) shows that the direct-sound decay curve is determined by the residual free-path distribution. Here, we establish the relationship between this general expression and the previously used exponential approximation, and then describe the decay of the total acoustic energy under complete wall absorption.

Normalizing the direct-sound energy by its value when the source stops, let $G_d(t)=E_d(t)/E_d(0)$. Then

$$
\begin{aligned}
G_d(t) &= \frac{1}{l_d}\int_{ct}^{\infty}\overline{F}_R(x)\,dx\,,\\
G_d(0) &= 1\,,\ \left.\frac{d\,G_d(t)}{dt}\right|_{t=0^+} = -\frac{c}{l_d}\,.
\end{aligned}
\qquad \text{(IV-D1)}
$$

The initial slope follows from the residual free path being positive with probability one. Thus, the mean residual free path $l_d$ determines both the energy when the source stops in Eq. (IV-C4) and the initial slope of the normalized decay curve. The full decay curve, however, requires the survival function $\overline{F}_R$.

A single exponential over all times requires an exponential residual survival function. The general curve in Eq. (IV-D1) is not restricted to this form.

The following direct-sound decay formula from the earlier studies [Hanyu2024; arXiv_V1] is an approximation to the energy decay curve:

$$E_d(t)\approx\frac{W\,l_d}{Vc}\exp\left(-ct/l_d\right). \qquad \text{(IV-D2)}$$

This approximation preserves the energy when the source stops and initial normalized slope, but not necessarily the full decay curve. Section VII compares the decay curves.

Under complete wall absorption, $\alpha=1$ and hence $q=0$, so the reflected component in Eq. (IV-C1) vanishes. Consequently,

$$F(t)=F_d(t)=\overline{F}_R(ct)\,,\,E_r(t)=0\,,\,E(t)=E_d(t). \qquad \text{(IV-D3)}$$

For $W>0$ and $l_d>0$, the energy when the source stops is $E(0)=W\,l_d/(Vc)>0$. Thus, even when complete wall absorption eliminates reflected sound, direct sound persists until its first wall collision [Hanyu2024; Hanyu2025; arXiv_V1]. In this case, the decay of the total acoustic energy is obtained solely from the residual free-path distribution, without using either the independent-renewal approximation or the exponential approximation.

## V. ENERGY LIFETIME AND MICROSCOPIC–MACROSCOPIC CONSISTENCY

### A. Mean acoustic energy lifetime and the previously developed macroscopic model

Acoustic energy emitted by a source propagates as direct sound until its first wall collision. The portion that is not absorbed at that collision continues to propagate after reflection. The mean lifetime of this entire process can be obtained from the initial direct-sound propagation time and the subsequent propagation times accumulated before absorption. Here, only wall absorption is considered, with $0<\alpha\le 1$.

Let $T_{\text{abs}}$ denote the time until a sound particle is absorbed according to the absorption rule specified in Sec. II A. Since the mean survival fraction defined in Sec. II C is $F(t)=Pr\{T_{\text{abs}}>t\}$, the mean acoustic energy lifetime $\tau_{imp}$ is

$$\tau_{imp}=E[T_{\text{abs}}]=\int_0^\infty F(t)\,dt. \tag{V-A1}$$

This is the integral of the survival fraction of acoustic energy supplied at time zero. It is distinct from the integral of the normalized decay curve after a source operating at constant acoustic power stops emitting sound.

Let $N_{\text{abs}}$ denote the order of the wall collision at which absorption occurs. The total propagation distance before absorption, $X_{\text{abs}}$, is

$$X_{\text{abs}}=c\,T_{\text{abs}}=R+\sum_{j=1}^{N_{\text{abs}}-1} L_j. \tag{V-A2}$$

Every sound particle travels the initial residual free path $R$. To travel the free path $L_j$ after the $j$th collision, however, it must survive the preceding $j$ collisions, which occurs with probability $q^j$. Using the independence of the absorption decisions from the geometrical trajectory specified in Sec. II A, the mean absorption free path $l_a$, defined in the earlier studies [Hanyu2024; Hanyu2025; arXiv_V1] as the mean total propagation distance before absorption, is obtained as

$$l_a=E[X_{\text{abs}}]=l_d+\sum_{j=1}^{\infty} q^j E[L_j],\ \tau_{imp}=\frac{l_a}{c}. \tag{V-A3}$$

The mean is assumed to be finite. Equation (V-A3) retains both correlations among the free paths and differences in their means with reflection order; it does not require the independent-renewal approximation.

If $E[L_j]=l$ for every reflection order, summing the geometric series gives

$$l_a=l_d+\frac{q}{\alpha}l,\ \tau_{imp}=\frac{1}{c}\left(l_d+\frac{q}{\alpha}l\right). \tag{V-A4}$$

This expression derives the mean absorption free-path formula of the earlier studies [Hanyu2024; Hanyu2025; arXiv_V1] from the expected propagation distances for each reflection order. The first term represents the direct-sound contribution up to the first wall collision, and the second represents the reflected-sound contribution accumulated before absorption. Under complete wall absorption, the second term vanishes, leaving a mean lifetime of $l_d/c$.

Equation (V-A4) requires a common mean free path $l$ at every reflection order, but does not require independence among the free paths. This condition is satisfied under the independent-renewal approximation in Sec. II D, but does not hold automatically for a general geometrical propagation process. When it is not satisfied, Eq. (V-A3) is used.

This mean lifetime provides a probabilistic interpretation of the decay rate in the previously developed macroscopic model [arXiv_V1]. In that model, the decay rate due to wall absorption was expressed as $c/l_a$, using the mean absorption free path $l_a$. Denoting this rate by $\lambda_{macro}$, Eq. (V-A3) gives

$$\lambda_{macro}=\frac{c}{l_a}=\frac{1}{\tau_{imp}}. \tag{V-A5}$$

Thus, the previously introduced macroscopic decay rate corresponds to the reciprocal of the mean acoustic energy lifetime. When $E\left[L_j\right]=l$ at every reflection order, Eq. (V-A4) yields

$$\lambda_{macro}=\frac{c}{l}\frac{\alpha}{1-\left(1-l_d/l\right)\alpha}, \tag{V-A6}$$

recovering the general macroscopic expression reported previously [arXiv_V1]. The purpose here is to relate the mean absorption free path underlying that decay rate to the mean lifetime of a sound particle, rather than to introduce a new definition of the rate.

This correspondence also appears in the steady-state energy. Equation (II-C4) of Sec. II C and Eq. (V-A1) give the energy density when the source stops for a source operating at constant acoustic power $W$:

$$E(0)=\frac{W}{V}\tau_{imp}=\frac{W}{V\lambda_{macro}}. \tag{V-A7}$$

The macroscopic model based on the mean lifetime therefore reproduces the acoustic energy stored in the room at steady state.

Let $E_{macro}(t)$ denote the energy density in the macroscopic model after the source stops. The previously developed energy equation [arXiv_V1] gives

$$\begin{aligned}\frac{dE_{macro}(t)}{dt} &= -\lambda_{macro}E_{macro}(t),\\ E_{macro}(t) &= E(0)\exp\left(-t/\tau_{imp}\right).\end{aligned} \tag{V-A8}$$

In addition to the energy when the source stops, this exponential curve preserves the initial slope of the normalized decay curve. Differentiating the general expression $E(t)=(W/V)\int_t^{\infty}F(\tau)\,d\tau$ and using $F(0)=1$ yields

$$\left.\frac{d}{dt}\frac{E(t)}{E(0)}\right|_{t=0^+}=-\frac{1}{\tau_{imp}}=-\lambda_{macro}, \tag{V-A9}$$

which agrees with the initial slope of Eq. (V-A8).

The macroscopic exponential therefore preserves mean lifetime, steady-state energy, and the initial normalized slope. These agreements do not require the full decay curve to be exponential.

### B. Long-time asymptotic decay rate and second-order microscopic–macroscopic consistency

We next obtain the long-time microscopic decay rate under the independent-renewal approximation for $0<\alpha<1$. Its low-absorption expansion refines the earlier macroscopic correspondence [arXiv_V1] using the second-moment relation in Sec. III.

Let $H(t)$ denote the mean acoustic energy survival fraction for a sound particle that begins propagation immediately after a wall collision. Separating trajectories according to whether the first wall-to-wall propagation continues beyond time $t$ or ends at an earlier collision gives

$$H(t)=\overline{F}_L(ct)+q\int_0^t\psi_L(\tau)H(t-\tau)\,d\tau. \tag{V-B1}$$

The second term includes the survival factor $q$ at the next collision and uses the fact that the same propagation process begins again after that collision. This renewal relation assumes independent and identically distributed wall-to-wall free paths.

Denoting the Laplace transform with respect to time by a hat, Eq. (V-B1) becomes

$$\hat{H}(s)=\frac{1-\hat{\psi}_L(s)}{s\left[1-q\,\hat{\psi}_L(s)\right]},\hat{\psi}_L(s)=E\left[\exp(-sL/c)\right]. \quad \text{(V-B2)}$$

The positive decay rate $\lambda_{tail}$ governing the long-time exponential decay corresponds to the zero of the denominator at $s=-\lambda_{tail}$ and satisfies

$$q\,E\left[\exp\left(\lambda_{tail}L/c\right)\right]=1. \quad \text{(V-B3)}$$

We assume that this positive root exists and that the required exponential moments are finite in its neighborhood. If the corresponding exponential moment of the initial residual free path is also finite and the initial propagation does not introduce a more slowly decaying tail, the same long-time decay rate applies to the delayed renewal process initiated within the room. The analysis below is restricted to this regime.

Equations (V-B1)–(V-B3) describe the long-time decay of the microscopic model under the independent-renewal approximation. Equation (V-B3) shows that its decay rate depends on the free-path distribution, rather than on the mean free path alone. At low absorption coefficients, however, the first two moments suffice to determine the expansion through second order. Introducing the dimensionless rate $z=\lambda_{tail}l/c$ and expanding Eq. (V-B3) near $\alpha=0$ gives

$$(1-\alpha)\left[1+z+\frac{E\left[L^2\right]}{2l^2}z^2+O\left(z^3\right)\right]=1. \quad \text{(V-B4)}$$

Substituting $z=\alpha+a_2\alpha^2+O\left(\alpha^3\right)$, comparing coefficients, and using $E\left[L^2\right]=2l\,l_d$, obtained from Eq. (III-B3), yields

$$\frac{\lambda_{tail}l}{c}=\alpha+\left(1-\frac{l_d}{l}\right)\alpha^2+O\left(\alpha^3\right). \quad \text{(V-B5)}$$

The mean residual free path $l_d$ thus determines the second-order coefficient of the long-time decay rate through the second moment of the wall-to-wall free path.

Expanding the macroscopic rate in Eq. (V-A6) gives

$$\frac{\lambda_{macro}l}{c}=\frac{\alpha}{1-\left(1-l_d/l\right)\alpha}=\alpha+\left(1-\frac{l_d}{l}\right)\alpha^2+O\left(\alpha^3\right). \quad \text{(V-B6)}$$

Consequently, under the independent-renewal approximation with common free-path statistics,

$$\lambda_{tail}-\lambda_{macro}=O\left(\frac{c}{l}\alpha^3\right). \quad \text{(V-B7)}$$

The earlier model used $l_d/c$ as the wall-to-wall propagation-time standard deviation [arXiv_V1], giving agreement at first order but differing second-order coefficients.

Using the wall-to-wall variance relation $Var(L)=2l\,l_d-l^2$ to specify propagation-time variance gives exact second-order agreement under the independent-renewal approximation with a common $l_d$.

This is an agreement of expansions, not an identity between mean lifetime and long-time rate. Higher free-path moments enter at third and higher orders.

For correlated propagation, Eq. (V-B1) does not apply directly, and the long-time rate depends on joint free-path statistics. The general mean-lifetime expression, Eq. (V-A3), does not require independence.

### C. Relationship among mean energy lifetime, asymptotic decay rate, and reverberation time

Rate-based reverberation formulas assume exponential decay. We distinguish their predictions from the 60-dB crossing of a general curve. A rate measured over a finite decay interval is also distinct from the mathematical long-time rate; that measurement issue is not examined here.

Equations (II-C5) and (II-C6) of Sec. II C determine the 60-dB crossing of the theoretical decay curve through

$$G(t)=\frac{E(t)}{E(0)}=\frac{1}{\tau_{imp}}\int_t^{\infty} F(\tau)\,d\tau\,, G(T_{60})=10^{-6}. \quad \text{(V-C1)}$$

The previously developed macroscopic model [arXiv_V1], by contrast, uses an exponential curve whose time constant is the mean lifetime, giving

$$T_{60}^{(macro)}=\frac{6\ln 10}{\lambda_{macro}}=6\ln 10\,\tau_{imp}. \quad \text{(V-C2)}$$

Similarly, the reverberation time converted from the long-time asymptotic decay rate of the microscopic model is

$$T_{60}^{(tail)}=\frac{6\ln 10}{\lambda_{tail}}. \quad \text{(V-C3)}$$

These values coincide when the decay curve is a single exponential. Since their agreement is not guaranteed for a general decay curve, the present study uses the crossing in Eq. (V-C1) as a reference for evaluating predictions based on decay rates. This distinction does not invalidate the engineering conversion of decay rate into reverberation time.

The correspondence between free-path models and classical reverberation formulas can also be established through Eq. (V-C3). Under the independent-renewal approximation, if the wall-to-wall free path takes the constant value $l$ with probability one, the relationship in Sec. III gives $l_d = l/2$. Equation (V-B3) then yields

$$\lambda_{tail}=-\frac{c}{l}\ln(1-\alpha). \quad \text{(V-C4)}$$

Substituting this rate into Eq. (V-C3), with $l=4V/S$, gives Eyring’s formula [Eyring1930]. This correspondence concerns the long-time asymptotic rate and does not require the entire decay curve, including the initial residual free path and discrete collision times, to be exactly a single exponential.

If the wall-to-wall free path instead follows an exponential distribution with mean $l$, then $l_d = l$, and the stationary residual free path follows the same exponential distribution. In this independent-renewal model, the collision count is a Poisson process with rate $c/l$, giving

$$F(t)=\exp(-c\alpha t/l)\,, \lambda_{tail}=\lambda_{macro}=\frac{c\alpha}{l}. \quad \text{(V-C5)}$$

The normalized decay curve after the source stops is then the same exponential function, and Eqs. (V-C1)–(V-C3) agree, yielding Sabine's formula [Sabine1922]. These results show how the classical formulas arise from the respective free-path models; they do not imply that actual room free-path distributions coincide with these models.

The microscopic model of the revised reverberation theory does not restrict the free path to a constant value or an exponential distribution. Instead, it uses a distribution that reflects both the mean free path $l$ and the mean residual free path $l_d$. Its long-time asymptotic decay rate is determined by Eq. (V-B3), and at low absorption coefficients,

$$\lambda_{tail} = \frac{c}{l}\left[\alpha + \left(1 - \frac{l_d}{l}\right)\alpha^2 + O\left(\alpha^3\right)\right]. \tag{V-C6}$$

Using the engineering approximation $l_d / l = 1 - 1/e$ from the earlier studies [Hanyu2024; arXiv_V1] gives

$$\lambda_{tail} = \frac{c}{l}\left[\alpha + \frac{1}{e}\alpha^2 + O\left(\alpha^3\right)\right]. \tag{V-C7}$$

This is the expansion of the microscopic decay rate of the revised theory, and it agrees through second order with the macroscopic rate using the same engineering approximation. A specific expression for the rate is obtained in the next section by specifying the free-path distribution.

All three models share the first-order term $c\alpha / l$, but their second-order coefficients differ: zero for the Sabine-type model, $1/2$ for the Eyring-type model, and $1/e$ for the engineering approximation of the revised theory. Under the independent-renewal approximation, these differences arise from the variance of the wall-to-wall free path.

In the constant-free-path model corresponding to Eyring's rate, all sound particles have the same intercollision interval, and the free-path variance is zero. When free paths vary, some trajectories undergo fewer wall collisions over the same elapsed time. Energy along those trajectories experiences fewer absorption events and is therefore more likely to survive. This effect slows the long-time decay. Indeed, the second-order coefficient in Eq. (V-C6) can be written as

$$1 - \frac{l_d}{l} = \frac{1}{2} - \frac{Var\left(L\right)}{2l^2}. \tag{V-C8}$$

Because a nonconstant free path has positive variance, this coefficient is smaller than the value $1/2$ of the constant-free-path model. Thus, in the low-absorption expansion, neglecting free-path variability contributes to overestimating the long-time decay rate and underestimating the reverberation time converted from that rate.

The exponential-free-path model corresponding to Sabine's rate, by contrast, assumes the specific variance $Var\left(L\right) = l^2$. If the free-path statistics used in the model have a smaller variance, the second-order coefficient in Eq. (V-C8) is positive, giving a larger decay rate than the Sabine-type model at low absorption coefficients. The engineering approximation of the revised theory gives $Var\left(L\right)/l^2 = 1 - 2/e$, placing it between the constant-free-path and exponential-free-path models.

Thus, $l_d$ connects the mean lifetime and microscopic rate through common free-path statistics, providing a statistical basis for the earlier theory [Hanyu2024; Hanyu2025; arXiv_V1]. The engineering ratio $l_d / l = 1 - 1/e$ is not universal, and correlations are neglected in the renewal model; Sec. VII evaluates the resulting predictions.

## VI. GAMMA-DISTRIBUTED MICROSCOPIC MODEL AND ENGINEERING APPROXIMATIONS

### A. Gamma-distributed microscopic model based on the first two free-path moments

Beyond the first two moments specified by $l$ and $l_d$, construction of the full decay curve requires a free-path distribution.

The normal approximations in the earlier microscopic models [Hanyu2025; arXiv_V1] assign probability to negative times. We retain their reflection-order convolution construction but use a positive gamma wall-to-wall free path with the moments from Sec. III, replacing the earlier propagation-time standard-deviation assumption $l_d / c$. The initial path uses the stationary residual distribution derived from the same free-path law.

Gamma approximations for flight times between successive boundary reflections have precedents in reverberation theory, as reviewed by Polack [Polack2025]. Here, the gamma approximation is combined with the moment relation in Sec. III and the stationary residual distribution derived from the same wall-to-wall free-path statistics.

Using the dimensionless variance $v$ defined in Sec. III C,

$$v = \frac{Var(L)}{l^2} = \frac{2 l_d}{l} - 1 . \qquad \text{(VI-A1)}$$

When the free path has nonzero variability ($v > 0$), the shape parameter $k$ and scale parameter $\theta$ of a gamma distribution with mean $l$ and variance $l^2 v$ are

$$k = \frac{1}{v}, \theta = lv = 2 l_d - l . \qquad \text{(VI-A2)}$$

Approximating the wall-to-wall free-path density by

$$p_L(x) = \frac{x^{k-1} \exp(-x/\theta)}{\Gamma(k) \theta^k}, x > 0 , \qquad \text{(VI-A3)}$$

gives $E[L] = k\theta = l$ and $Var(L) = k\theta^2 = l^2 v$, preserving the moment relationship established in Sec. III. Here, $\Gamma$ denotes the gamma function. The constant-free-path model with $v = 0$ corresponds to the limit $v \to 0$.

For the initial residual free path, a separate gamma distribution is not specified independently. Instead, the stationary residual distribution obtained from Eq. (III-A5) is used. Expressing the gamma survival function through the regularized upper incomplete gamma function $Q$ gives

$$\overline{F}_L(x) = Q(k, x/\theta), p_R(r) = \frac{1}{l} Q(k, r/\theta). \qquad \text{(VI-A4)}$$

Here, $Q(a,z) = \Gamma(a,z) / \Gamma(a)$, with $\Gamma(a,z) = \int_z^{\infty} u^{a-1} e^{-u} du$. The residual distribution in Eq. (VI-A4) is generally not a gamma distribution, but its mean is $l_d$.

Equations (IV-B1)–(IV-B4) are used to convert the distance distributions into time distributions and construct the sequential convolutions under the independent-renewal approximation. Since the sum of independent gamma variables with a common scale parameter is also gamma distributed, for $n \geq 2$,

$$f_n(t) = \left(\psi_R * g_{(n-1)k, \theta/c}\right)(t), f_1(t) = \psi_R(t). \qquad \text{(VI-A5)}$$

Here, $g_{a,b}$ denotes a gamma probability density with shape parameter $a$ and time scale parameter $b$, and $\psi_R(t) = c\, p_R(ct)$. The reflection-order collision-time density $f_n$ therefore generally does not reduce to a single gamma density, because it includes convolution with the initial residual propagation time.

Equation (VI-A5), followed by Eq. (IV-B5), gives $Q_n(t)$. The decomposition and backward integration in Sec. IV C then yield direct, reflected, and total energy decay.

Within the independent-renewal framework, preserving the first two moments also preserves mean lifetime and the second-order low-absorption decay coefficient. Higher moments follow the gamma approximation, and serial correlations are neglected. The construction therefore does not assert exact reflection-order arrival-time distributions for geometric trajectories.

### B. Analytical decay rate and macroscopic correspondence

Substitution of the gamma distribution into Eq. (V-B3) gives its long-time rate under the independent-renewal approximation, for $v > 0$ and $0 < \alpha < 1$.

The exponential moment of the gamma distribution in Eq. (VI-A3) is

$$E\left[\exp(\lambda L / c)\right] = (1 - \theta\lambda / c)^{-k}, \lambda < c / \theta. \tag{VI-B1}$$

Substitution into Eq. (V-B3) gives $q\left(1 - \theta\, \lambda_{tail} / c\right)^{-k} = 1$. Using $k = 1/v$ and $\theta = lv$ from Eq. (VI-A2), the long-time asymptotic decay rate of the gamma-distributed microscopic model is obtained as

$$\lambda_{tail} = \frac{c}{lv}\left[1 - (1 - \alpha)^{v}\right]. \tag{VI-B2}$$

This positive root is smaller than $c / \theta$, and the corresponding exponential moment of the stationary residual distribution in the preceding subsection is also finite. The conditions specified in Sec. V B are therefore satisfied.

The relation between gamma-distributed free-flight times and reverberation decay rates has also been examined by Polack [Polack2025]. The present formulation makes the role of the mean residual free path explicit through the second-moment relation established in Sec. III and its connection to the macroscopic model.

Converting this rate into a reverberation time using Eq. (V-C3) gives

$$T_{60}^{(tail)} = \frac{6 \ln 10\, lv}{c\left[1 - (1 - \alpha)^{v}\right]}. \tag{VI-B3}$$

The decay curve itself is obtained through the convolution and backward integration described in the preceding subsection, whereas the reverberation time converted from the long-time decay rate can be evaluated using this compact expression.

The previously developed macroscopic formula [arXiv_V1], using the same $l_d$, can be expressed through Eq. (V-A6) and $l_d / l = (1 + v) / 2$ as

$$\lambda_{macro} = \frac{c}{l} \frac{\alpha}{1 - (1 - v)\alpha / 2}. \tag{VI-B4}$$

Expanding Eq. (VI-B2) at low absorption coefficients yields

$$\frac{\lambda_{tail}\, l}{c} = \alpha + \frac{1 - v}{2}\alpha^2 + \frac{(1 - v)(2 - v)}{6}\alpha^3 + O(\alpha^4). \tag{VI-B5}$$

The macroscopic second-order coefficient is likewise $(1-\nu)/2$, but its third-order coefficient $(1-\nu)^2/4$ generally differs. The rates therefore agree through second order, not over the full absorption range.

Equation (VI-B2) also clarifies the relationship between free-path variability and the classical formulas. As $\nu \to 0$, the gamma distribution approaches a constant free path, and $\lambda_{tail} \to -(c/l)\ln(1-\alpha)$, reproducing the Eyring-type decay rate. At $\nu=1$, the gamma distribution becomes exponential, and $\lambda_{tail}=c\alpha/l$, reproducing the Sabine-type rate. For $0<\nu<1$, the gamma microscopic decay rate lies between the Sabine-type and Eyring-type rates for the same $l$ and $\alpha$. This relationship follows from the free-path variance rather than from an empirical interpolation between the two reverberation formulas.

Under complete wall absorption, $\alpha=1$, reflected sound vanishes, and the total energy is governed by direct-sound decay. In this case, the general expression in Sec. IV D is used; Eq. (VI-B3) is not directly identified with the actual reverberation time.

### C. Engineering approximations and model construction using common parameters

Because $l=4V/S$ is geometric while $l_d$ depends on additional free-path statistics, compact engineering formulas require an approximation for their ratio.

The mean residual free-path ratio $l_d/l$ and the dimensionless variance $\nu$ are related by

$$l=\frac{4V}{S},\ \nu=2\frac{l_d}{l}-1. \qquad \text{(VI-C1)}$$

The gamma model requires $l_d/l>1/2$. Equations (V-C2) and (VI-B3) give the reverberation time of the macroscopic model and that converted from the long-time decay rate of the gamma microscopic model as

$$\begin{aligned} T_{60}^{(macro)} &= \frac{24\ln 10\,V}{cS}\frac{1-(1-l_d/l)\alpha}{\alpha} \approx \frac{55.3\,V}{cS}\frac{1-(1-l_d/l)\alpha}{\alpha}, \\ T_{60}^{(tail)} &= \frac{24\ln 10\,V}{cS}\frac{2l_d/l-1}{1-(1-\alpha)^{2l_d/l-1}} \approx \frac{55.3\,V}{cS}\frac{2l_d/l-1}{1-(1-\alpha)^{2l_d/l-1}}. \end{aligned} \qquad \text{(VI-C2)}$$

The former corresponds to the previously developed macroscopic formula based on the mean lifetime [arXiv_V1], whereas the latter follows from the gamma free-path model used in the present study. The macroscopic expression is used for $0<\alpha\le 1$, and the expression based on the long-time decay rate for $0<\alpha<1$.

We examine two engineering approximations: $l_d/l=1-1/e$, inherited from the earlier studies, and the alternative $l_d/l=2/3$. The former is adopted as the primary engineering approximation [Hanyu2024; arXiv_V1]. It is based on a geometrical cutoff model using the mean free path as the reference length and is not an exact stationary statistical value for arbitrary room geometries. The present study retains its physical motivation and simplicity while applying the same approximation to the second free-path moment. Equation (VI-C2) becomes

$$\begin{aligned} T_{60}^{(macro)} &= \frac{24\ln 10\,V}{cS}\frac{1-\alpha/e}{\alpha} \approx \frac{55.3\,V}{cS}\frac{1-\alpha/e}{\alpha}, \\ T_{60}^{(tail)} &= \frac{24\ln 10\,V}{cS}\frac{1-2/e}{1-(1-\alpha)^{1-2/e}} \approx \frac{55.3\,V}{cS}\frac{1-2/e}{1-(1-\alpha)^{1-2/e}}. \end{aligned} \qquad \text{(VI-C3)}$$

The first expression recovers the earlier macroscopic formula; the gamma expression uses $\nu=1-2/e$ and shares its low-absorption expansion through second order.

Next, we consider the alternative $l_d/l=2/3$. This value is relatively close to the geometrical mean residual free-path ratios of the cubic and rectangular rooms examined, and it gives $v=1/3$ and a gamma shape parameter $k=3$, allowing a simple model construction. In this case,

$$\begin{aligned} T_{60}^{(macro)} &= \frac{24\ln 10\, V}{cS}\frac{1-\alpha/3}{\alpha} \approx \frac{55.3\, V}{cS}\frac{1-\alpha/3}{\alpha}, \\ T_{60}^{(tail)} &= \frac{24\ln 10\, V}{3cS\left[1-(1-\alpha)^{1/3}\right]} \approx \frac{55.3\, V}{3cS\left[1-(1-\alpha)^{1/3}\right]}. \end{aligned} \tag{VI-C4}$$

This value is a candidate selected with reference to the free-path statistics of the geometries examined and the simplicity of the model; it is not a universal constant derived from Lambertian reflection conditions. Comparison with the primary approximation inherited from the earlier studies is used to assess the sensitivity of reverberation predictions to the mean residual free-path ratio.

For either ratio $l_d/l$, Eq. (VI-A2) supplies $k=1/(2l_d/l-1)$ and $\theta=l(2l_d/l-1)$; the initial residual density follows from the same gamma law. A geometry-specific $l_d$ can instead be used without a fixed ratio.

Common parameters permit comparison of mean lifetime, long-time rate, and decay curves without making them identical. For complete absorption, the residual distribution is used with Sec. IV D.

Three approximations remain distinct: independence neglects serial correlations, the gamma law represents the distribution while preserving two moments, and the fixed ratio $l_d/l$ supplies an engineering estimate. Their predictive consequences are examined next.

## VII. NUMERICAL VERIFICATION

### A. Ray-tracing simulation conditions and procedures

Sections VII A–D describe the three-room verification under spatially uniform initial conditions.
Section VII E presents a separate fixed-source and local-receiver test using T30.

Ray-tracing simulations were performed for three rectangular rooms to verify the statistical free-path relationships and reverberation decay models presented in the preceding sections. The room side-length ratios were 1:1:1, 10:4:3, and 15:4:2, with a common volume of $512\,m^3$. The corresponding side lengths were $8.000\,m$, $8.000\,m$, and $8.000\,m$; $16.219\,m$, $6.488\,m$, and $4.866\,m$; and $24.329\,m$, $6.488\,m$, and $3.244\,m$, respectively. This configuration allowed the effects of room geometry on free-path statistics and reverberation prediction to be compared at a fixed room volume.

The sound speed was $340\,m/s$, and air absorption was neglected. All walls had a scattering coefficient of unity, with diffuse reflections governed by Lambert's cosine law. The absorption coefficient $\alpha$ was uniform over all walls. Seven absorption coefficients, 0.05, 0.1, 0.2, 0.3, 0.4, 0.5, and 1, were considered, giving 21 combinations of room geometry and absorption coefficient. Engineering prediction accuracy was evaluated for $\alpha \le 0.5$, designated here as the practical absorption range. Complete absorption, $\alpha=1$, was considered to examine direct-sound decay and the physical characteristics of the models.

For each room, five runs with different random seeds were performed, each using $10^6$ sound particles. Initial positions were sampled uniformly within the room, and initial propagation directions were sampled isotropically. Wall intersections were determined geometrically, and a new propagation direction was generated at each collision. For each particle, the residual free path $R$, from its initial position to the first wall, and the subsequent 600 wall-to-wall free paths $L$ were recorded. Geometrical paths were generated independently of absorption losses, and the reflection losses corresponding to each absorption coefficient were applied to the same paths. Free-path statistics were therefore calculated without absorption weighting.

The distributions and moments of $R$ and $L$ were evaluated and compared with the mean free path $l = 4V/S$ and the stationary residual free-path relationships. The mean wall-to-wall free path was calculated separately for each reflection order to examine the influence of the initial state. Data from reflection order 5 onward were used for the principal distribution and moment estimates. Pearson correlation coefficients at lags of one, two, and three reflection intervals were also calculated to assess the interval-independence assumption underlying the renewal approximation. The first wall-to-wall interval following the initial wall collision was indexed as order 1. Free-path distributions and order-dependent means were averaged over the five runs; correlation coefficients were calculated separately for each run and then averaged.

For the energy-decay calculation, $N(t)$ denotes the number of wall collisions experienced by a particle up to time $t$. The energy survival function $F(t)$ was estimated by averaging the surviving particle weights, $(1-\alpha)^{N(t)}$, over the particle ensemble. A time increment of $0.25\,ms$ was used. The normalized energy decay after a steady source stops emitting sound was obtained by backward integration of $F(t)$:

$$E(t)/E(0) = \frac{\int_t^{t_{max}} F(u)\,du}{\int_0^{t_{max}} F(u)\,du} \tag{VII-A1}$$

The numerical integration limit $t_{max}$ was set to at least 1.5 times the reverberation time predicted by the revised macroscopic formula with the ratio $2/3$ for the slowest-decaying condition, $\alpha = 0.05$. The integration limits were $6.40\,s$, $5.70\,s$, and $4.77\,s$ for the rooms with side-length ratios of 1:1:1, 10:4:3, and 15:4:2, respectively. Under an exponential-decay approximation, these limits correspond to approximately $-90\,dB$, providing a sufficient margin for evaluating reverberation time at $-60\,dB$. All particle paths were confirmed to extend to the integration limit in every run.

The energy decay in each run was expressed as $10\log_{10}[E(t)/E(0)]$. The reverberation time $T_{60}$ was determined as the time at which the normalized energy reached $10^{-6}$, using linear interpolation between adjacent samples in decibels. Mean decay curves were obtained by averaging the normalized linear energies over the five runs before conversion to decibels. In contrast, statistics of reverberation time and relative prediction error were calculated from the values determined separately for each run, rather than from the mean decay curve.

The comparisons included the Sabine and Eyring formulas, the revised macroscopic formulas, and the gamma-distributed microscopic models. The revised macroscopic formulas used $1-1/e$ and $2/3$ for the ratio of the mean residual free path to the mean free path, $l_d/l$. The gamma-distributed microscopic models were evaluated using either geometry-derived free-path moments or these two fixed ratios. For the geometry-based gamma model, parameters were estimated from the free-path statistics of each run, and its predictions were compared with the ray-tracing results from that same run. Its mean decay curve was likewise obtained by averaging the normalized linear energies before conversion to decibels. For the gamma energy-decay calculations, the mean free path was set to 4V/S and the dimensionless variance was estimated from the wall-to-wall intervals of order 5 onward in each run. The geometry-based mean residual free path used in these curves was then specified by the corresponding moment relation. The independent stationary sample used for Table I was evaluated separately.

The relative reverberation-time error was defined as $100\left(T_{60,model}-T_{60,ray}\right)/T_{60,ray}$, with the ray-tracing result as the reference. For each condition, the mean and sample standard deviation of the relative errors over the five runs were calculated to evaluate prediction error and variability between runs. An error range of $\pm 5\,\%$ was adopted as an engineering reference for prediction accuracy in the practical absorption range. Under completely absorptive conditions, the evaluation focused on the shape of the direct-sound decay after the source stops and its finite duration imposed by room geometry, rather than on reverberation-time prediction accuracy.

### B. Free-path statistics and assessment of the renewal approximation

Figure 1 compares the wall-to-wall and residual densities, $L$ and $R$, with their gamma-based approximations, using distances normalized by $l=4V/S$. Simulated densities are averaged over five runs, and the gamma law uses the first two moments averaged over those runs.

The wall-to-wall free-path distributions exhibit distinct features associated with room geometry. A single prominent peak appears in the cube, whereas multiple peaks appear in the rooms with unequal side lengths. These distributions retain their geometry-dependent features even after normalization by the mean free path, indicating that the mean alone does not characterize the free-path statistics.

The gamma model represents geometry through the first two free-path moments. Figure 1 assesses this statistical representation using simulated moments; fixed-ratio engineering approximations are evaluated below through decay and reverberation-time predictions.

Residual densities are smoother and decrease with distance, consistent with their proportionality to the wall-to-wall upper-tail probability in Sec. III. The gamma-derived residual densities capture this trend and determine direct-sound decay through the backward integral in Sec. IV C.

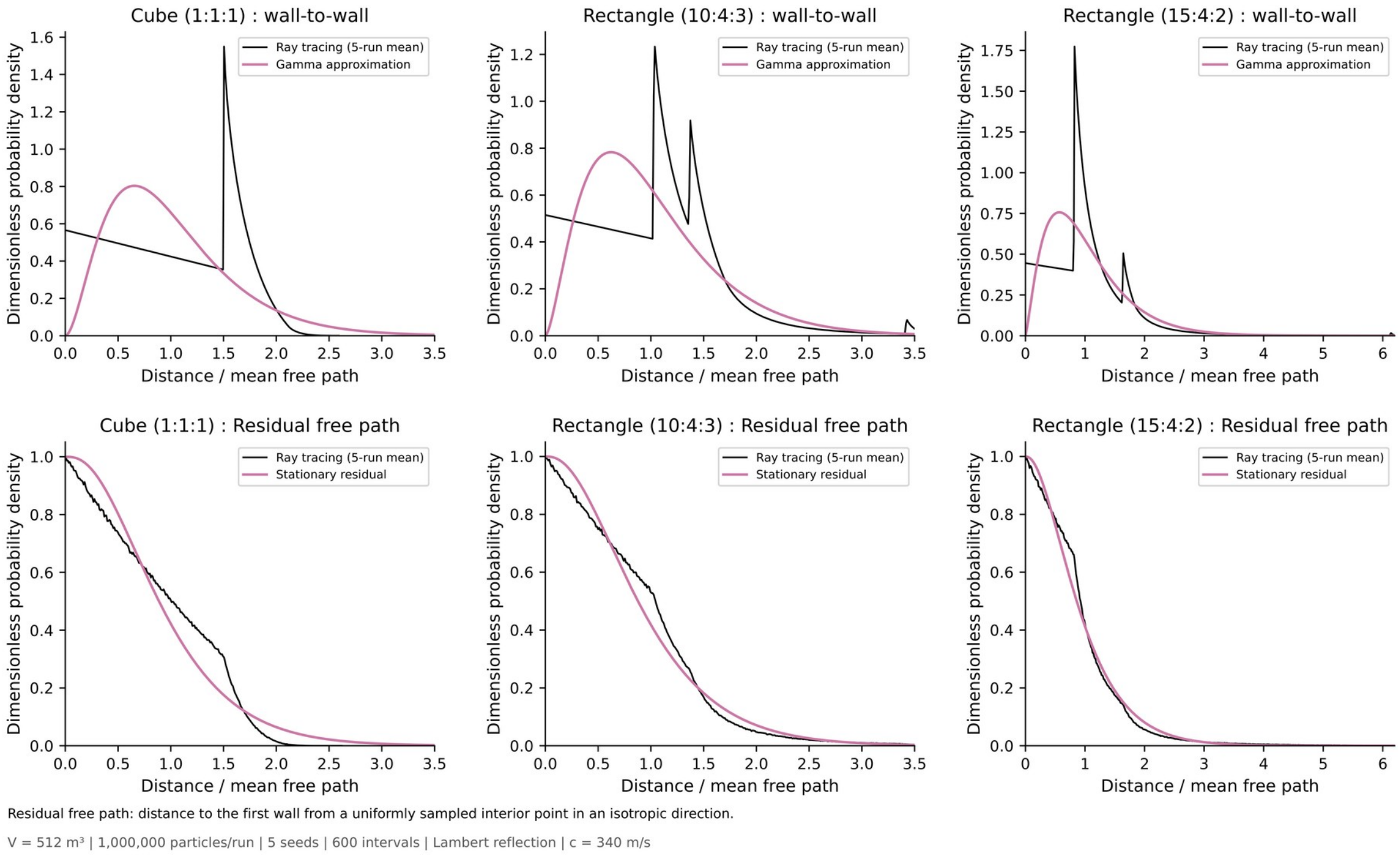

**FIG. 1. Free-path distributions and geometry-matched gamma approximations.**

Free-path distributions obtained by ray tracing and their gamma-model approximations in three rectangular rooms of equal volume (512 $m^3$). Columns correspond to side-length ratios of 1:1:1, 10:4:3, and 15:4:2, respectively. The upper row shows wall-to-wall free paths, and the lower row shows residual free paths, defined as the distances to the first wall from uniformly sampled interior points along isotropically sampled directions. Distances are normalized by the mean free path, 4V/S, and probability densities are expressed in dimensionless form. Simulated distributions show the mean of the probability densities from five runs with different random seeds. The gamma approximation is constructed from the first and second wall-to-wall free-path moments averaged over these runs, using intervals from order 5 onward; the corresponding stationary residual distribution is shown in the lower row. Each run used $10^6$ particles and 600 wall-to-wall intervals per particle.

The upper panels of Fig. 2 show the mean wall-to-wall free path at each reflection order, normalized by $l=4V/S$. Order 1 denotes the first wall-to-wall interval following the initial wall encounter. The plotted values are averages over five runs, and the vertical-axis limits are common to the three rooms. In all three rooms, the mean initially exceeds unity slightly and approaches unity as the reflection order increases. Most of the change occurs during the first few reflections. The influence of the initial state persists somewhat longer in the room with a side-length ratio of 15:4:2, but the means in all three rooms remain close to unity at the later orders shown. Thus, the first wall-to-wall intervals following initialization at uniformly distributed interior points exhibit small departures from the stationary mean.

The lower panels of Fig. 2 show Pearson correlation coefficients at lags of one, two, and three reflection intervals, evaluated using wall-to-wall free paths from order 5 onward. Each coefficient is calculated separately for each run and then averaged over the five runs. Positive correlations are observed between adjacent intervals in all three rooms, and their magnitudes decrease with increasing lag. These results indicate that successive free paths are not strictly independent, even after the order-dependent mean has become nearly constant. Although a new direction is sampled according to Lambert's cosine law at each wall encounter, the starting position of the next interval is determined by the endpoint of the preceding interval. The observed correlations are consistent with this geometrical dependence.

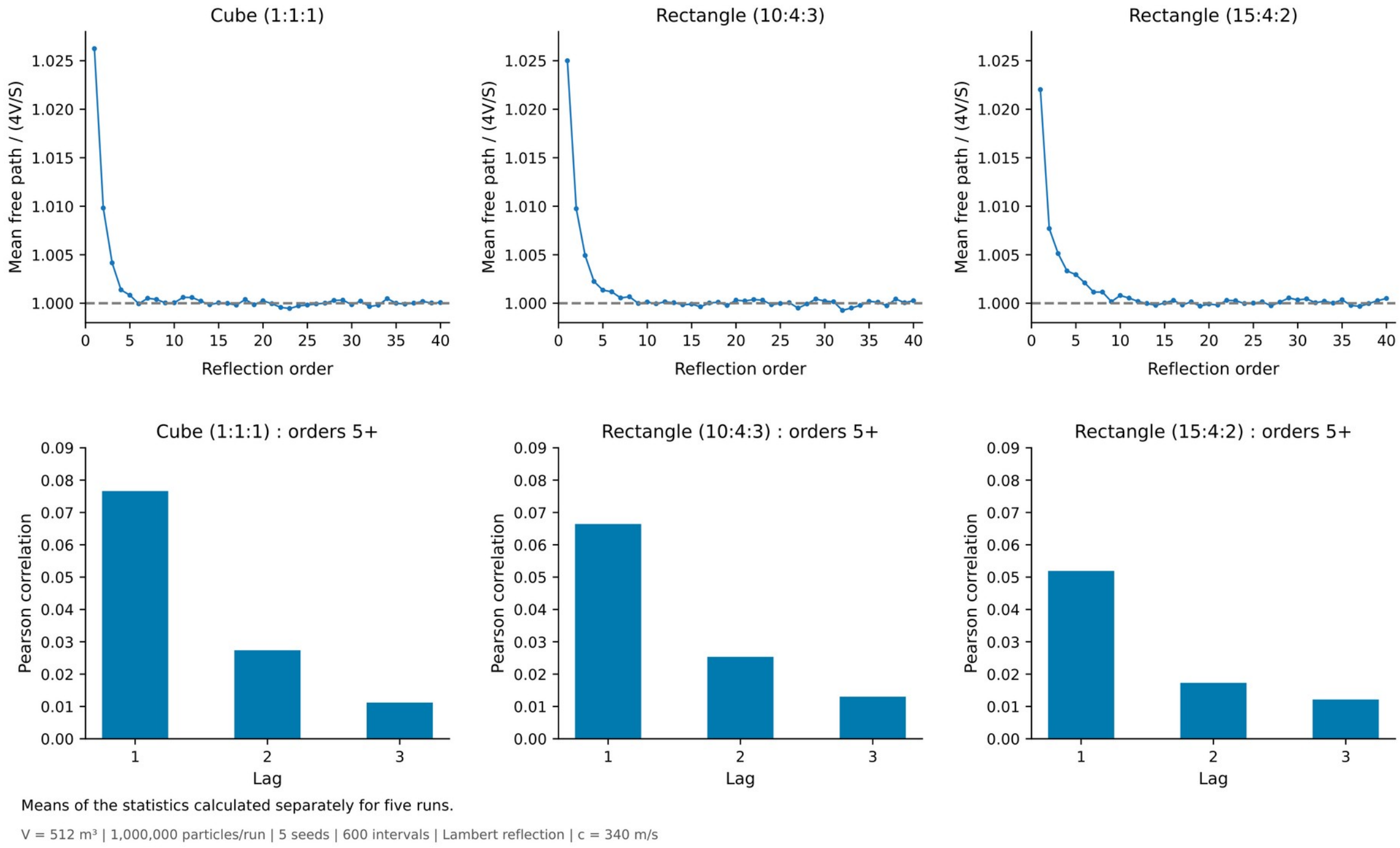

**FIG. 2. Early-order relaxation and observed correlations (lags 1–3).**

Reflection-order dependence of the mean wall-to-wall free path and serial correlations between successive free paths. Columns correspond to the three room geometries shown in Fig. 1. Upper panels show the mean free path at each reflection order up to order 40, normalized by 4V/S; dashed horizontal lines indicate unity. Order 1 denotes the first wall-to-wall interval following the initial wall encounter. Vertical-axis limits and tick intervals are common to all three upper panels. Lower panels show Pearson correlation coefficients at lags of one, two, and three reflection intervals, evaluated using wall-to-wall free paths from order 5 onward. Both rows show means of the statistics calculated separately for five runs with different random seeds. Each run used $10^6$ particles and 600 wall-to-wall intervals per particle, with Lambertian wall reflections.

An additional analysis of the reflection-order wall-collision time distributions showed that their temporal variance increased more rapidly than predicted from independent wall-to-wall intervals with the same stationary free-path variance. This difference was observed over both orders 5–20 and orders 20–200 and is consistent with a contribution from positive correlations between successive free paths. It does not invalidate the stationary residual-path moment relation, but shows that the marginal free-path moments alone do not fully determine the temporal variance of cumulative propagation times. Quantifying the contribution of correlations to this variance expansion and to reverberation decay remains a subject for future work. The increase rates were estimated by fitting straight lines with free intercepts to the variances obtained from 10-ms arrival-time histograms, using only reflection orders for which all particles were recorded. Binning uncertainty is distinct from the variability across runs.

The gamma approximation for individual free paths and the independent-renewal approximation therefore address distinct aspects of the model. The following comparisons assess their combined effectiveness for predicting energy decay and reverberation time. Agreement in these quantities does not, by itself, establish independence of successive free paths or exact agreement of reflection-order arrival-time distributions.

## C. Energy decay in the practical absorption range and under completely absorptive conditions

Figure 3 compares decay at $\alpha = 0.1, 0.3, 0.5$. Ray-tracing and geometry-based gamma curves use the five-run linear-energy averaging described in Sec. VII A.

The comparisons include the Sabine and Eyring formulas, the revised macroscopic formula using $l_d / l = 1 - 1/e$, and the gamma models using either free-path moments or fixed ratios. The curves obtained from the macroscopic formulas are represented by single-exponential decays corresponding to their respective reverberation times. The gamma-model curves are obtained by backward integration of the energy survival functions based on residual and wall-to-wall free-path statistics.

At $\alpha = 0.1$, the predictions of the different models are relatively close in all three rooms. As the absorption coefficient increases to 0.3 and 0.5, differences in decay slope become more evident. Under the conditions shown, the Sabine formula predicts slower decay than the ray-tracing simulations, whereas the Eyring formula predicts faster decay. The curves of the revised macroscopic formula and the gamma models generally lie between these two classical predictions and are closer to the ray-tracing results.

The ray-tracing curves are approximately linear on the decibel scale over the practical absorption conditions shown. The decay rate also varies with room geometry, even when the volume and absorption coefficient are held constant. For the three rooms considered here, the surface area increases in the order of side-length ratios 1:1:1, 10:4:3, and 15:4:2. The mean free path $l = 4V/S$ consequently decreases, increasing the average frequency of wall encounters. In addition to this basic geometrical effect, free-path variability and residual free-path statistics contribute to differences among the model predictions.

The gamma model using free-path moments gives decay curves close to the ray-tracing results across the three rooms. For the gamma models using fixed ratios, the differences vary with room geometry and absorption coefficient and become relatively apparent at $\alpha = 0.5$ in the room with a side-length ratio of 15:4:2. The statistical formulation of the decay model and the engineering approximation used to specify its input statistics from a small set of basic variables should therefore be assessed separately. The engineering prediction accuracy of the revised macroscopic formulas and the gamma models is compared in the next section using relative reverberation-time errors.

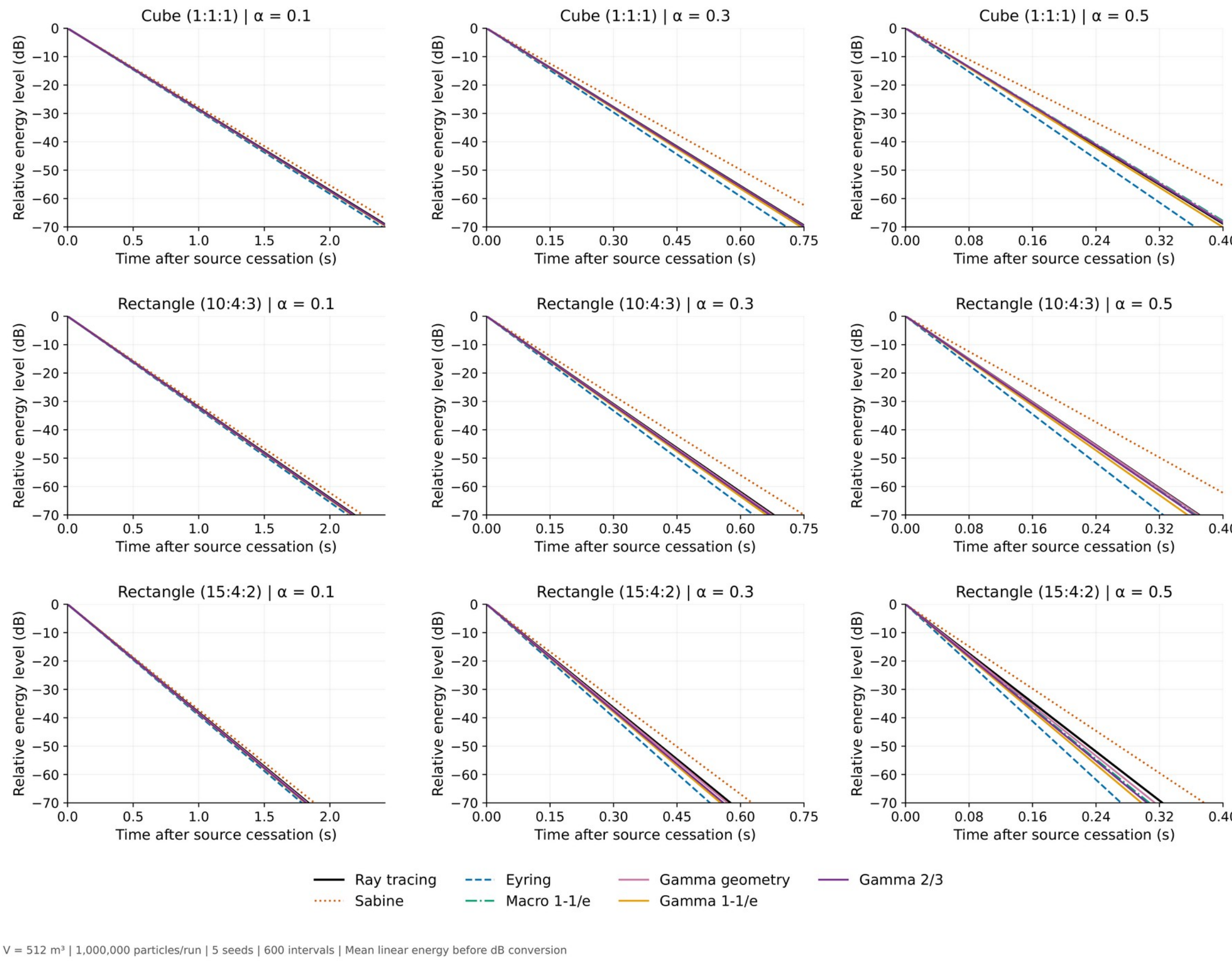

**FIG. 3. Decay after the source stops in the practical absorption range.**

Relative energy decay after the source stops in the practical absorption range. Rows correspond to room side-length ratios of 1:1:1, 10:4:3, and 15:4:2, and columns to uniform absorption coefficients of α = 0.1, 0.3, and 0.5. Ray-tracing results are compared with the Sabine and Eyring predictions, the revised macroscopic formula using the ratio 1 − 1/e, and the gamma models using geometry-derived parameters or the fixed ratios 1 − 1/e and 2/3. Energy levels are normalized to their values when the source stops. The time-axis range is identical across the three rooms within each column. Ray-tracing curves show the mean of the normalized linear-energy curves from five runs with different random seeds, converted to decibels after averaging. The geometry-based gamma curves are averaged in the same way using the model parameters estimated in each run. Each run used $10^6$ particles and 600 wall-to-wall intervals per particle.

Figure 4 compares direct-sound decay under completely absorptive conditions, $\alpha = 1$, including Sabine's prediction as a reference alongside the revised macroscopic and gamma models. The same five-run linear-energy averaging is used.

Despite the absence of reflections, all three simulated rooms show decay after the source stops: propagating direct sound persists until it reaches the first wall.

Immediately after the source stops, the Sabine formula gives a shallower decay slope than the ray-tracing results, whereas the revised models using the mean residual free path $l_d$ show closer agreement with the simulated slopes. Under completely absorptive conditions, the initial slope of the normalized energy predicted by the Sabine formula is $-c/l$. In contrast, the theoretical initial slope of direct-sound decay based on the residual free path is $-c/l_d$. The corresponding initial slopes in decibels are $-10c/(l \ln 10)$ and $-10c/(l_d \ln 10)$, respectively.

Table II quantifies this initial-slope comparison. The geometry-based prediction uses sampled wall-to-wall free-path moments, whereas the fixed-ratio models require only room volume and surface area.

At later times, the ray-tracing curves exhibit a rapid decrease near the end of the decay. This behavior arises because the room dimensions are finite, imposing an upper bound on the travel time before direct sound reaches a wall. The vertical dotted lines indicate the room diagonal divided by the sound speed, which provides a geometrical upper bound on the remaining travel time of direct sound in a rectangular room. In contrast, the statistical models based on gamma distributions or exponential decay do not terminate at a finite time.

This comparison tests the physical features of direct-sound decay, not engineering prediction of reverberation time under completely absorptive conditions or the exact termination of decay in an individual room.

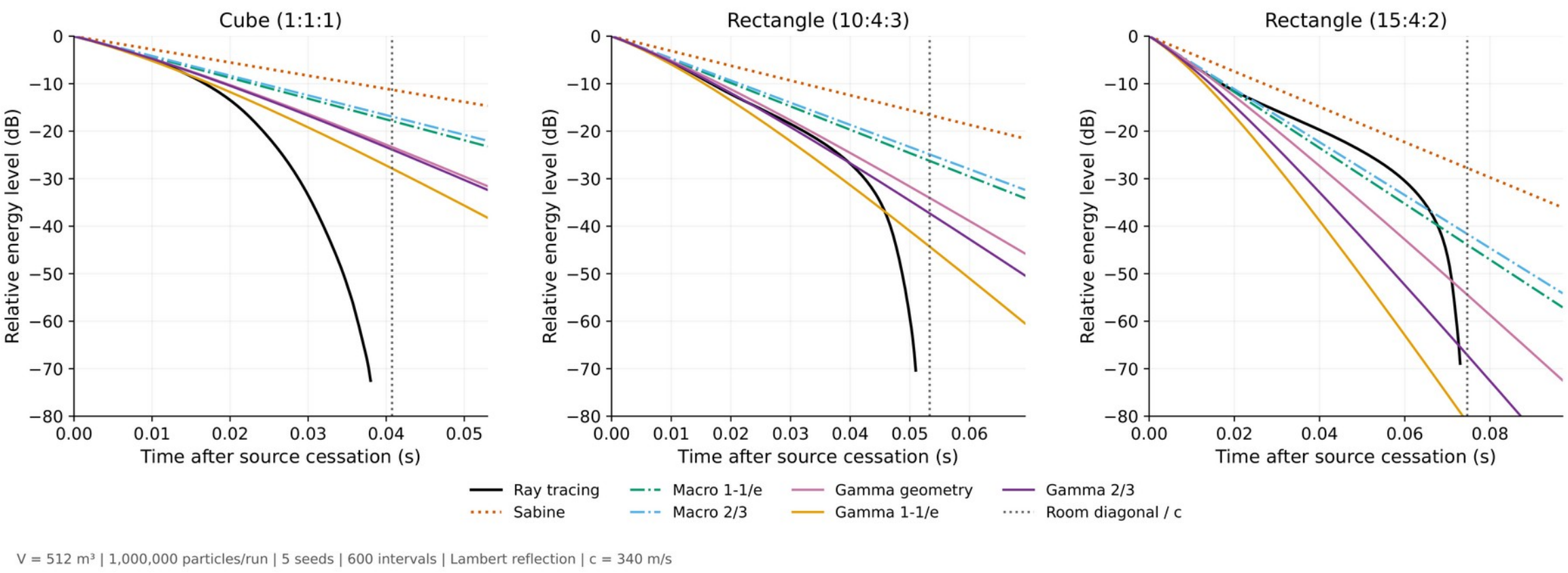

**FIG. 4. Direct-sound decay under completely absorptive conditions (α = 1).**

Direct-sound energy decay after the source stops under complete boundary absorption (α = 1). Columns correspond to room side-length ratios of 1:1:1, 10:4:3, and 15:4:2. Ray-tracing results are compared with the revised macroscopic formulas and gamma models; the Sabine formula applied at α = 1 is included as a reference. The comparison highlights the initial slopes associated with the mean wall-to-wall and residual free paths, as well as the rapid decrease near the end of the simulated decay in finite rooms. Vertical dotted lines indicate the room diagonal divided by the sound speed, a geometrical upper bound on the remaining travel time to the first wall. Energy levels are normalized to their values when the source stops. Ray-tracing and geometry-based gamma curves show normalized linear energies averaged over five runs and then converted to decibels; the geometry-based model parameters are estimated separately in each run. Each run used $10^6$ particles and 600 wall-to-wall intervals per particle.

The steady-state direct-sound energy provides a complementary check on the role of the mean residual free path. Under completely absorptive conditions, this energy was evaluated by numerically integrating the fraction of particles that had not yet undergone their first wall collision and multiplying the integral by the source sound power. This calculation did not use an estimated mean residual free path as an input. The survival fraction was integrated using the trapezoidal rule with a time step of approximately 0.147 ms, continuing until it reached zero. The resulting integrals differed by less than 0.0024% from the time-bin occupancy integrals in all 15 runs.

The prediction error is expressed as an energy-level difference,

$$\Delta L_U = 10 \log_{10}\left(U_{pred} / U_{d,ray}\right) \quad \text{(VII-C1)}$$

where $U_{pred}$ is the predicted steady-state room-integrated energy, given by either $W l_d / c$ or $W l / c$, and $U_{d,ray}$ is the steady-state direct-sound energy obtained by integrating the zeroth-order particle survival fraction and multiplying by $W$. The factor 10 is used because the ratio compares energies rather than pressure amplitudes. A positive value indicates overprediction relative to ray tracing, a negative value indicates underprediction, and zero indicates agreement.

Table I compares the resulting energy with the predictions $W l_d / c$ and $W l / c$. When both length scales were estimated from the same independently sampled stationary free-path ensemble, the prediction $W l_d / c$ agreed with the ray-tracing result to within 0.003 dB in the five-run mean for each room. This agreement provides a numerical consistency check on the residual-path moment relation under the prescribed spatially uniform and isotropic initial conditions. In contrast, $W l / c$, corresponding to the Sabine steady-state energy at complete absorption, overpredicted the energy by approximately 1.45–1.73 dB.

The fixed-ratio approximations were also evaluated using $l = 4V / S$. The approximation $l_d = l(1 - 1/e)$ yielded level differences from −0.54 to −0.27 dB, whereas $l_d = 2l/3$ yielded differences from −0.31 to −0.04 dB. Thus, both approximations reduced the discrepancy relative to the Sabine prediction without requiring sampled free-path statistics, although their accuracy depended on room geometry. These comparisons concern the room-integrated direct-sound energy under completely absorptive conditions and are distinct from the reverberation-time assessment in the practical absorption range.

TABLE I. Differences between predicted and ray-tracing estimates of steady-state direct-sound energy under completely absorptive conditions.

All entries are energy-level differences $\Delta L_U$ in decibels, calculated as $10 \log_{10}(U_{pred} / U_{d,ray})$, where $U_{pred}$ is the energy predicted by the expression specified in each row and $U_{d,ray}$ is the ray-tracing estimate of steady-state direct-sound energy. Positive and negative values indicate overprediction and underprediction, respectively. For the first two rows, $l = E[L]$ and $l_d = E[L^2] / (2E[L])$ were estimated from the same independently sampled stationary Lambertian free-path ensemble. The remaining rows use geometric inputs and the indicated fixed ratios. Values are averages of the level differences calculated separately for five runs; sample standard deviations were below 0.006 dB in all cases. Each run used $10^6$ particles with uniformly distributed initial positions and isotropic initial directions. Here, $c$ is the sound speed.

| **Prediction** | **Cube 1:1:1** | **Rectangular room 10:4:3** | **Rectangular room 15:4:2** |
|---|---|---|---|
| $W l_d / c$, sampled free-path moments | +0.001 | −0.002 | −0.003 |
| $W l / c$, sampled mean free path | +1.726 | +1.620 | +1.451 |
| $W l / c,\ l = 4V / S$ | +1.723 | +1.619 | +1.450 |
| $W l_d / c,\ l = 4V / S,\ l_d = l(1 - 1/e)$ | −0.269 | −0.373 | −0.542 |
| $W l_d / c,\ l = 4V / S,\ l_d = 2l/3$ | −0.038 | −0.142 | −0.311 |

To quantify the initial slope, each normalized direct-sound energy curve was converted to decibels and fitted separately over ranges from 0 to −5, −2, −1, and −0.5 dB. Linear and quadratic least-squares fits used the original 0.25-ms samples and a freely fitted intercept. The coefficient of time in the quadratic fit estimates the derivative when the source stops while accounting for local curvature. Table II reports the magnitude of this derivative from the 0 to −1 dB fit, using 10, 9, and 8 samples for the three rooms, respectively. Means and sample standard deviations were calculated from the five individual runs, rather than by fitting the mean decay curve.

For the geometry-based initial-slope prediction, the mean residual free path was obtained from the second moment and mean of the independently sampled stationary Lambertian wall-to-wall free paths used in Table I. The mean initial residual path was not used as a prediction input. This choice differs from the order-5-onward statistics used for the geometry-based gamma curves in Fig. 4 and provides a separate sampling check of the same moment relation. The other predictions use the geometrical mean free path and the indicated fixed ratios. The revised macroscopic and gamma models predict identical initial slopes when supplied with the same mean residual free path.

As the fitting interval was shortened, the linear-fit slopes approached the residual-path prediction. Quadratic fits reduced the influence of curvature: over 0 to −1 dB, the absolute five-run mean relative discrepancies from the stationary-free-path prediction were 0.076%, 0.112%, and 0.079% for the 1:1:1, 10:4:3, and 15:4:2 rooms, respectively. Fits over 0 to −2 dB gave discrepancies below 0.5% in all three rooms. The 0 to −0.5 dB fits gave similar estimates but contained only four or five samples. These interval checks support close agreement without interpreting the between-run standard deviations as total numerical uncertainty.

The Sabine prediction gives a shallower initial slope in all three rooms. The fixed-ratio revised predictions are closer to the simulations, although their deviations depend on geometry. This comparison concerns the derivative when the source stops, not a finite-interval decay time or agreement over the entire curve. Because the decay is obtained by backward integration of direct-sound survival, its initial slope and the steady-state energy in Table I are mathematically related. The comparison therefore checks the connection to separately sampled wall-to-wall statistics rather than providing an independent validation of the full decay model.

**TABLE II. Initial direct-sound decay slopes under completely absorptive conditions.**

Entries are magnitudes of the initial slopes in dB/s. Ray-tracing estimates are the means ± one sample standard deviation of quadratic-fit derivatives when the source stops, evaluated separately over 0 to −1 dB for five runs. The geometry-based prediction uses the stationary wall-to-wall free-path moments sampled separately from the initial residual paths; its between-run standard deviations are 0.07, 0.43, and 0.63 dB/s, respectively. Fixed-ratio and Sabine predictions use the geometrical mean free path and do not vary between runs. A common mean residual free path gives the same initial slope for the revised macroscopic and gamma models.

| Method | 1:1:1 | 10:4:3 | 15:4:2 |
|---|---|---|---|
| Ray tracing<br>Quadratic fit | 411.25 ± 0.16 | 451.25 ± 0.37 | 519.03 ± 0.53 |
| Revised theory<br>Geometry-based | 411.56 | 451.75 | 519.44 |
| Revised theory<br>Ratio 1 − 1/e | 437.99 | 492.08 | 588.10 |
| Revised theory<br>Ratio 2/3 | 415.29 | 466.58 | 557.62 |
| Sabine | 276.86 | 311.05 | 371.75 |

### D. Reverberation-time prediction accuracy in the practical absorption range

Figure 5 compares relative reverberation-time errors for $\alpha \leq 0.5$: 18 combinations of three rooms and six absorption coefficients. Five-run means and sample standard deviations are evaluated as specified in Sec. VII A.

The shaded ±5% bands are an engineering reference. Here, maximum absolute error means the largest absolute five-run mean error among the 18 conditions.

The Sabine formula overestimates reverberation time, whereas the Eyring formula underestimates it. For both formulas, the absolute error increases with absorption coefficient. The maximum absolute errors are 24.30% for the Sabine formula and 16.05% for the Eyring formula. The mean relative errors fall within ±5% in 6 and 8 of the 18 conditions, respectively. Errors also depend on room geometry at a given absorption coefficient: the Sabine overestimation is largest in the cube, whereas the Eyring underestimation is largest in the room with a side-length ratio of 15:4:2.

With $l_d / l = 1 - 1/e$, the revised macroscopic formula has a maximum absolute error of 5.03%, with 17 of 18 mean errors within ±5%. The exception is $\alpha = 0.5$ in the 15:4:2 room (−5.03%), slightly outside the reference band. With $l_d / l = 2/3$, the maximum is 3.58%, and all 18 conditions are within ±5%.

Both fixed ratios substantially improve on the classical formulas using only room volume, surface area, and absorption coefficient. Although $2/3$ gives a smaller maximum error here, the central engineering result is the useful accuracy retained with these few inputs.

The gamma model using free-path moments has a maximum absolute error of 2.84%, with all 18 mean errors within ±5%. Parameters are determined from each run's free-path statistics, not fitted to reverberation time. This supports prediction of reverberation time in the tested rooms, but not exact reproduction of serial correlations or reflection-order temporal variance. The separate effects of correlations and distributional approximations on the remaining errors are unresolved.

For the gamma models using fixed ratios, the maximum absolute errors are 8.10% for $1 - 1/e$ and 5.95% for $2/3$. The mean relative errors fall within ±5% in 16 and 17 of the 18 conditions, respectively. In particular, reverberation-time underestimation increases with absorption coefficient in the room with a side-length ratio of 15:4:2. Even when the same fixed ratio is used, the gamma model does not always yield a smaller error than the revised macroscopic formula. This result indicates that the effect of approximating the free-path statistics by a fixed ratio also depends on the decay model in which that approximation is used.

Across all models and conditions, the sample standard deviation is at most approximately 0.17 percentage points, so many error bars are hidden by the markers. Seed-to-seed variability is small relative to the main model differences, but does not represent total model uncertainty.

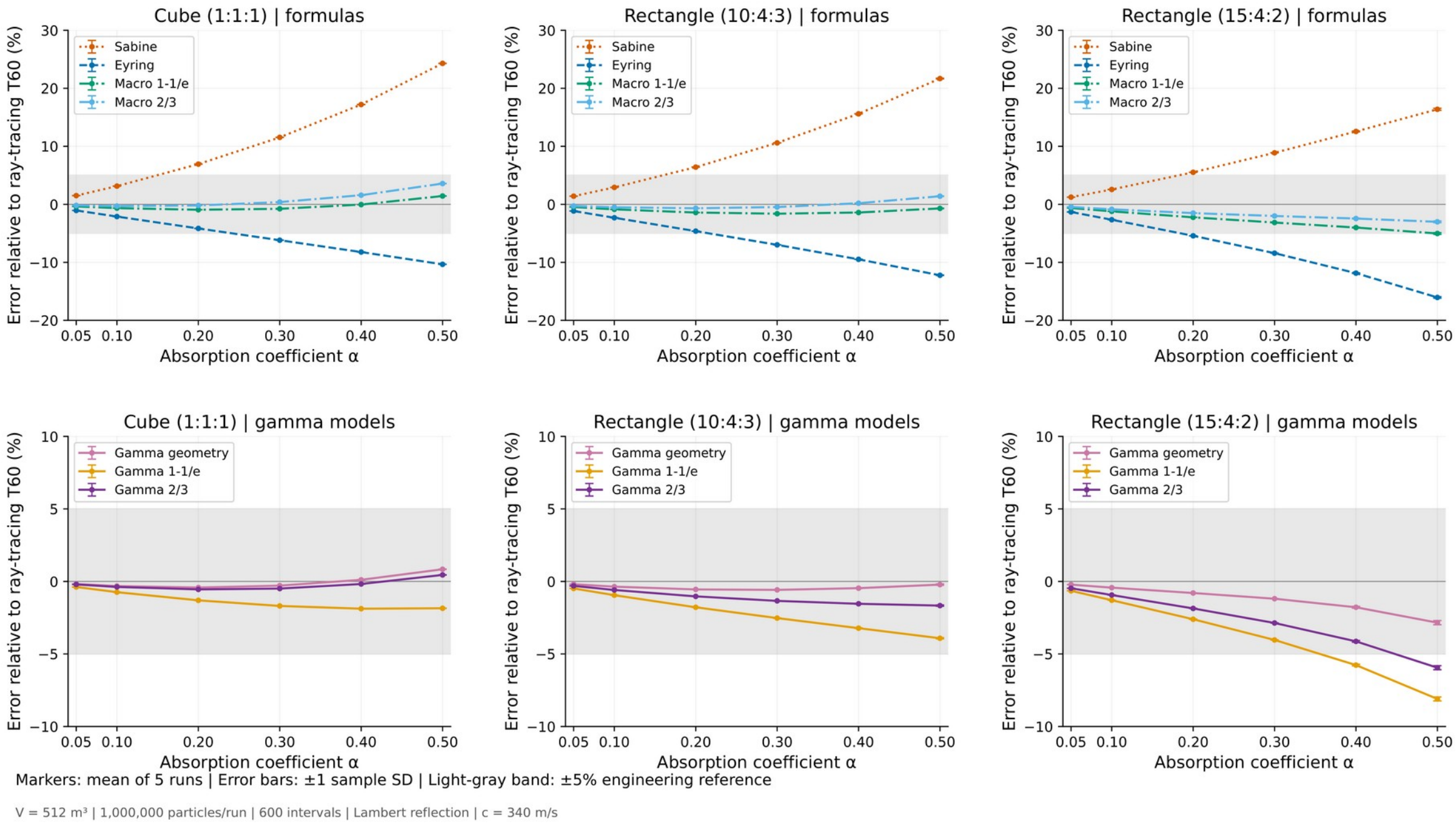

**FIG. 5. Reverberation-time prediction errors (α ≤ 0.5).**

Relative errors in the predicted reverberation time $T_{60}$ for uniform absorption coefficients α ≤ 0.5. Columns correspond to the three room geometries. Upper panels compare the Sabine, Eyring, and revised macroscopic formulas; lower panels compare the gamma models using geometry-derived parameters or the fixed ratios 1 − 1/e and 2/3. Errors are defined as $100(T_{60},\text{model} - T_{60},\text{ray})/T_{60},\text{ray}$, where the ray-tracing result is the reference. The light-gray bands indicate ±5%, adopted as an engineering reference for prediction accuracy. Vertical-axis limits are common to all panels within each row. Markers show the mean of the relative errors calculated separately for five runs with different random seeds; error bars indicate ±1 sample standard deviation, not confidence intervals. Many error bars are smaller than the markers and are therefore not visible at the scale shown. Each run used $10^6$ particles and 600 wall-to-wall intervals per particle. For the geometry-based gamma model, the free-path parameters were estimated separately in each run.

## E. Supplementary verification for fixed sources and local receivers

The theoretical target remains the source-position-averaged room energy defined in Sec. II. This supplementary test examines whether its predictions are also useful engineering approximations for T30 at fixed source and receiver positions. T30 is obtained by least-squares fitting the normalized decay in decibels from −5 to −35 dB and extrapolating the fitted slope to a 60-dB decay. It is distinct from the initial-value-to−60-dB crossing time used in Secs. VII A–D.

A separate rectangular room measuring 5 m × 4 m × 3 m was used, with sound speed 340 m/s, no air absorption, Lambertian reflection with scattering coefficient unity, and uniform wall absorption coefficients of 0.1, 0.3, and 0.5. Relative to a room corner, the omnidirectional sources S1, S2, and S3 were located at (2.5, 2.0, 1.5), (0.5, 0.5, 0.5), and (1.0, 2.8, 1.2) m. Spherical receivers R1, R2, and R3 were centered at (1.0, 1.2, 1.5), (4.0, 2.8, 1.5), and (2.5, 2.0, 0.5) m, respectively, with radius 0.25 m. The three sources, three receivers, and three absorption coefficients give 27 source–receiver conditions.

Five runs with different random seeds were performed for each source and for the spatially uniform, isotropic initial-condition reference. Each run used $10^6$ particles and 600 wall-to-wall intervals after the first wall collision. Absorption losses were applied to common geometrical paths. Receiver energy was accumulated from absorption-weighted residence times inside each sphere, divided by its volume, and backward-integrated to obtain decay after the steady source stops. Direct and reflected contributions were both retained. The output time step was 0.5 ms and the integration limit was 2 s. Each curve was normalized to its own value when the source stops. T30 was evaluated separately for each run; statistics were not obtained by fitting an averaged curve or by pooling different positions.

Figure 6 uses the room-integrated result for spatially uniform initial conditions as the common reference. Differences are calculated from the five-run means as 100 times the ratio of the tested T30 to the reference T30, minus 100. The reference means were 1.006324, 0.310555, and 0.167579 s for absorption coefficients of 0.1, 0.3, and 0.5, respectively. Fixed-source whole-room T30 differed from this reference by at most 0.035%. Receiver differences ranged from −0.037% to +0.047%, −0.059% to +0.157%, and −0.176% to +1.001%, respectively. Thus, within these conditions, replacing the averaged initial source position with a fixed source and evaluating local energy produced only small changes in T30.

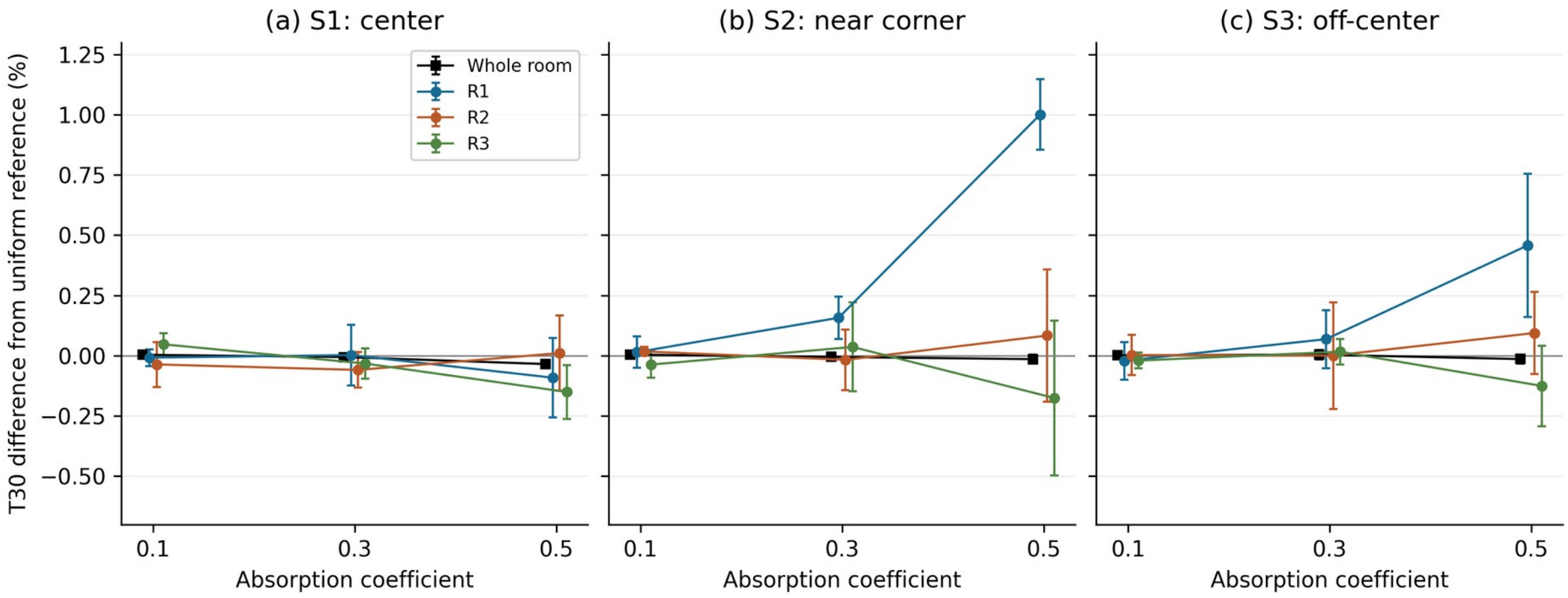

FIG. 6. Source- and receiver-position dependence of T30.

Differences from the five-run mean whole-room T30 for spatially uniform and isotropic initial conditions in the same 5 m × 4 m × 3 m room. Panels correspond to sources S1–S3. Squares denote fixed-source whole-room results; circles denote receivers R1–R3. Markers show differences between means, and error bars show ± one sample standard deviation of the tested T30, divided by the common reference mean and multiplied by 100. They are not standard deviations of paired differences or confidence intervals and do not include uncertainty in the reference mean. The reference coefficients of variation were below 0.027%. Symbols are slightly offset horizontally for readability; vertical-axis limits are common to all panels.

Table III compares model predictions directly with the receiver T30 values. Sabine, Eyring, and revised macroscopic predictions were represented by their single-exponential decay curves. For each gamma model, T30 was evaluated over the same −5 to −35 dB interval of its predicted decay. The mean free path was 4V/S. Geometry-derived gamma parameters used the dimensionless wall-to-wall free-path variance estimated from order 5 onward separately in each run; they were not fitted to receiver decay. The stationary residual distribution was retained, without introducing source- or receiver-specific model corrections. Fixed-ratio models used the same ratios as in Sec. VII D. For each run and condition, the relative error was 100(T30,model − T30,ray)/T30,ray; these errors were then averaged over the five runs.

The maximum absolute mean errors were 1.22% and 3.36% for the revised macroscopic formulas with ratios 1 − 1/e and 2/3, respectively. Both formulas were within the ±5% engineering reference for all 27 conditions. The corresponding maxima were 24.04% for Sabine and 11.57% for Eyring. All three gamma variants were also within ±5%. Geometry-derived gamma inputs contain additional simulated information, so their accuracy should be distinguished from that of formulas requiring only basic room variables.

TABLE III. T30 prediction errors for the 27 fixed-source–receiver conditions.

Entries summarize the five-run mean relative errors (%). The error range and maximum absolute error are taken across 27 conditions, not across individual runs. The final column gives the number of conditions whose mean error lies within ±5%.

| **Model** | **Mean error range (%)** | **Max. absolute (%)** | **Within ±5%** |
|---|---|---|---|
| Sabine | +3.05 to +24.04 | 24.04 | 9/27 |
| Eyring | -11.57 to -2.11 | 11.57 | 9/27 |
| Revised macro: 1 − 1/e | -1.09 to +1.22 | 1.22 | 27/27 |
| Revised macro: 2/3 | -0.39 to +3.36 | 3.36 | 27/27 |
| Gamma: geometry-derived | -0.48 to +0.85 | 0.85 | 27/27 |
| Gamma: 1 − 1/e | -3.22 to -0.75 | 3.22 | 27/27 |
| Gamma: 2/3 | -0.96 to +0.21 | 0.96 | 27/27 |

The receiver T30 coefficients of variation were at most 0.322% across the five runs. For S3, additional five-run tests changed the receiver radius to 0.15 or 0.35 m, or doubled the particle count to $2 \times 10^6$ at radius 0.25 m. Across the receivers and absorption coefficients, the largest absolute changes in mean T30 from the baseline were 0.059%, 0.058%, and 0.069%, respectively. No receiver condition triggered a missing-signal or tail warning; the final 10% of the integration interval contributed less than $2.13 \times 10^{-11}$ of the accumulated energy. These checks support numerical stability for the tested settings; sensitivity tests were limited to S3, and between-run variability does not represent total numerical or model uncertainty.

These results support the engineering applicability of the revised formulas to fixed-source–receiver T30 in the tested room. They do not change the theoretical averaging assumptions, imply agreement over the entire decay curve, or establish applicability to arbitrary geometries, nonuniform absorption, or weaker scattering. The conclusions concern this additional room and the specified absorption range.

## VIII. CONCLUSIONS

This study gave the previously proposed revised reverberation theory [Hanyu2024; Hanyu2025; arXiv_V1] a common statistical basis for direct-sound persistence, reflected-sound propagation, and room energy decay.

Complete absorption reveals its physical starting point: direct sound produces decay after the source stops even without reflections. This explains why eliminating reflected sound does not imply instantaneous loss of room energy. In the simulations, initial slopes were closer to models using the mean residual free path than to Sabine's model using the wall-to-wall mean, supporting this physical interpretation.

The mean residual free path also characterizes variability after reflection. The earlier model assumed $(l_d/c)^2$ for temporal variance [Hanyu2025; arXiv_V1]. Here, the integral-geometric relation $Var(L)=2l\,l_d-l^2$ determines the wall-to-wall distance variance. Division by the squared sound speed gives the interval-time variance, which determines temporal variance expansion under independent renewal. This connects direct sound to the statistical description of reflected sound.

Using common free-path statistics, the microscopic long-time rate and macroscopic rate agree exactly through second order in absorption under independent renewal, extending the earlier approximate correspondence. The macroscopic rate is the reciprocal mean energy lifetime; the result does not identify this quantity with the long-time rate at arbitrary absorption.

A gamma wall-to-wall distribution with its consistent residual distribution removes negative arrival times while preserving the first two moments. Geometry-derived parameters yielded accurate reverberation times in the three rooms. Observed serial correlations and excess temporal variance expansion nevertheless limit the independent-renewal approximation; their quantitative effect on decay remains to be resolved.

Fixed-ratio approximations make the macroscopic formula calculable from volume, surface area, and absorption coefficient. In the tested practical absorption range, they substantially reduced the errors of the Sabine and Eyring formulas. Thus, useful prediction accuracy was maintained despite representing reverberation through only a few inputs.

The supplementary fixed-source and local-receiver test provides a limited connection to spatially resolved acoustic evaluation. In the additional room with uniform absorption and Lambertian reflection, both revised macroscopic formulas predicted receiver T30 with mean errors within ±5% for all 27 conditions. This supports engineering applicability under the tested conditions while retaining the source-position-averaged theoretical formulation.

The resulting framework connects the physical insight obtained from direct-sound reverberation to the entire reverberation process, with mean and residual free paths linking its statistical foundation to engineering prediction.